\documentclass[runningheads, envcountsame]{llncs}
\usepackage[utf8]{inputenc}
\usepackage[T1]{fontenc}
\usepackage{graphicx}
\usepackage[margin=1in]{geometry}
\usepackage{hyperref}
\usepackage{amsfonts}
\usepackage{mathtools}
\usepackage{bbm}
\usepackage{subcaption}
\usepackage{tikz}
\usepackage{algorithm}
\usepackage{algpseudocode}
\usepackage[capitalise, nameinlink, noabbrev]{cleveref} 

\DeclarePairedDelimiterX{\inp}[2]{\langle}{\rangle}{#1, #2}
\newcommand{\norm}[1]{\left\lVert#1\right\rVert}
\newcommand{\sort}[1]{#1_{\downarrow}}
\newcommand{\decleq}{\leq_{dec}}

\DeclareMathOperator*{\argmax}{arg\,max}
\DeclareMathOperator*{\argmin}{arg\,min}
\begin{document}
\title{Faster Primal-Dual Convergence for Min-Max Resource Sharing and Stronger Bounds via Local Weak Duality}
\titlerunning{Faster Primal-Dual Convergence for Min-Max Resource Sharing}
%
\author{Daniel Blankenburg}
\authorrunning{D. Blankenburg}
%
\institute{Research Institute for Discrete Mathematics, University of Bonn, Bonn, Germany\\
\email{blankenburg@or.uni-bonn.de}}
\maketitle              
\begin{abstract}
We revisit the (block-angular) min-max resource sharing problem, which is a well-known generalization of fractional packing and the maximum concurrent flow problem. It consists of finding an $\ell_{\infty}$-minimal element in a Minkowski sum $\mathcal{X}= \sum_{C \in \mathcal{C}} X_C$ of non-empty closed convex sets $X_C \subseteq \mathbb{R}^{\mathcal{R}}_{\geq 0}$, where $\mathcal{C}$ and $\mathcal{R}$ are finite sets. We assume that an oracle for approximate linear minimization over $X_C$ is given. \par
In this setting, the currently fastest known FPTAS is due to Müller, Radke, and Vygen \cite{min_max_resource_sharing_mrv}. For $\delta \in (0,1]$, it computes a $\sigma(1+\delta)$-approximately optimal solution using  $\mathcal{O}((|\mathcal{C}|+|\mathcal{R}|)\log |\mathcal{R}| (\delta^{-2} + \log \log |\mathcal{R}|))$ oracle calls, where $\sigma$ is the approximation ratio of the oracle.\par
We describe an extension of their algorithm and improve on previous results in various ways. Our FPTAS, which, as previous approaches, is based on the multiplicative weight update method, computes close to optimal primal and dual solutions using $\mathcal{O}\left(\frac{|\mathcal{C}|+ |\mathcal{R}|}{\delta^2} \log |\mathcal{R}|\right)$ oracle calls.  \par
We prove that our running time is optimal under certain assumptions, implying that no warm-start analysis of the algorithm is possible.\par 
A major novelty of our analysis is the concept of local weak duality, which illustrates that the algorithm optimizes (close to) independent parts of the instance separately. Interestingly, this implies that the computed solution is not only approximately $\ell_{\infty}$-minimal, but among such solutions, also its second-highest entry is approximately minimal.
We prove that this statement cannot be extended to the third-highest entry.

\keywords{Resource sharing \and Dantzig-Wolfe-type algorithms \and Decreasing minimization}
\end{abstract}

\section{Introduction}
\subsection{Problem description}
Dividing a limited set of \emph{resources} among \emph{customers} is an overarching theme of numerous problems in discrete and continuous optimization. A common formulation of such problems is known as \emph{min-max resource sharing} in the literature. In this work, we consider the \emph{block-angular min-max resource sharing problem} as it was first studied by Grigoriadis and Khachiyan \cite{grigoriadis_many_blocks}. The problem consists of choosing a feasible resource allocation for every customer, such that the maximum accumulated resource usage is minimized. Formally, it can be described as follows: \par 
There is a finite set of customers $\mathcal{C} $ and a finite set of resources $\mathcal{R}$. We denote their sizes by $n := |\mathcal{C}|$ and $m:= |\mathcal{R}|$. For each customer $C \in \mathcal{C}$, there is a non-empty closed convex set $X_C\subseteq \mathbb{R}^m_{\geq 0}$ of \emph{feasible resource allocations}, also referred to as \emph{block}. The set of \emph{feasible solutions} of the (block-angular) min-max resource sharing problem is given as the Minkowski-sum $ \mathcal{X} := \sum_{C \in \mathcal{C}} X_C$.  \par
Further, we assume that we are given, for some constant $\sigma \geq 1$, a $\sigma$-approximate \emph{block-solver}, which is an approximate linear minimization oracle for non-negative price vectors. It is specified by functions $f_C: \mathbb{R}_{\geq 0}^m \rightarrow X_C$  for all $C \in \mathcal{C}$ that satisfy
\[ \forall y \in \mathbb{R}^m_{\geq 0}: \qquad \inp{y}{f_C(y)} \leq \sigma opt_C(y), \]
where $opt_C(y) := \min_{x \in X_C} \inp{y}{x}$.
\begin{problem}[Block-angular min-max resource sharing problem]
\label{prob:resource_sharing}
With the notation above, the (block-angular) \emph{min-max resource sharing problem} is to compute resource allocations $x_C \in X_C$ for every customer $C \in \mathcal{C}$, such that $x := \sum_{C \in \mathcal{C}} x_C$ attains
\[ \lambda^* := \min_{x \in \mathcal{X}} \norm{x}_{\infty}.\]
\end{problem}
In the following, we abbreviate this problem as the \emph{resource sharing problem}. For shorter notation, we call $\mathcal{X}$ an instance of the resource sharing problem, meaning that we associate with $\mathcal{X}$ the oracle functions and the explicit decomposition into the blocks $X_C$. \par
In this work, we consider fully polynomial-time approximation schemes relative to $\sigma$. For $\delta > 0$ we construct a solution $x \in \mathcal{X}$ with $ \norm{x}_{\infty} \leq \sigma (1+\delta)\lambda^*$ within a number of oracle calls that is polynomial in $n,m$, and $\delta^{-1}$. \par
Algorithms that interact with the feasible region only via a linear minimization oracle are known as algorithms of the \emph{Dantzig-Wolfe type}. An iteration consists of choosing a price vector $y \in \mathbb{R}^m_{\geq 0}$ and querying the linear minimization oracle of a customer $C \in \mathcal{C}$ with $y$. The computed solution is a convex combination of the solutions returned from the oracle. At their core, Dantzig-Wolfe-type algorithms are primal-dual algorithms.
In the case of the resource sharing problem, the dual is to find $q \in \Delta_m:= \{p \in [0,1]^m: \norm{p}_1 = 1\}$, such that $\min_{x \in \mathcal{X}} \inp{q}{x} = \max_{p \in \Delta_m} \min_{x \in \mathcal{X}} \inp{p}{x}$. Strong duality is implied by von Neumann's minimax theorem:
\[ \max_{p \in \Delta_m} \min_{x \in \mathcal{X}} \inp{p}{x} = \min_{x \in \mathcal{X}} \max_{p \in \Delta_m} \inp{p}{x}= \min_{x \in \mathcal{X}} \norm{x}_{\infty} = \lambda^*.\]
\begin{problem}[Dual of the resource sharing problem]
Given an instance $\mathcal{X}$ of the resource sharing problem, find a $p \in \Delta_m$ such that 
\[ \min_{x \in \mathcal{X}} \inp{p}{x} = \lambda^*. \]
\end{problem}

\begin{figure}[t]
     \begin{subfigure}[b]{0.45\textwidth}
         \centering
         \includegraphics[width=0.5\textwidth]{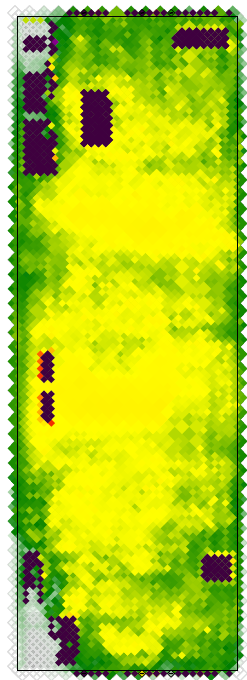}
         \caption{High congestion only in local hotspots.}
         \label{subfig:only_local_congestion}
     \end{subfigure}
     \hfill
   \begin{subfigure}[b]{0.45\textwidth}
         \centering
         \includegraphics[width=0.5\textwidth]{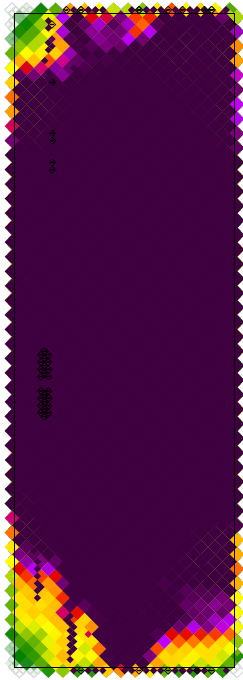}
         \caption{High congestion in almost every part.}
         \label{subfig:completely_congested}
     \end{subfigure}
        \caption{Comparison of two fractional routing results. Congestion of global routing tiles is visualized by color (lowest congestion in green, highest in violet). }
        \label{fig:routing_comparison}
\end{figure}
Our work is motivated  by applications to global routing in VLSI design, where a resource sharing formulation has proven successful in theory and practice \cite{vlsi_routing_application,global_routing_with_timing,min_max_resource_sharing_mrv}. There, highly simplified, the customers correspond to nets --- sets of terminals on the chip --- that have to be connected by a Steiner tree, resources to edges of the global routing graph, and feasible resource allocations to the convex hull of incidence vectors of Steiner trees. One seeks to find a collection of (fractional) Steiner trees that minimize the maximum edge overload. The block solvers can be implemented by approximate shortest Steiner tree algorithms. \cref{fig:routing_comparison} depicts two fractional routing results on an industrial microprocessor. The maximum edge congestion is the same in both cases, so both solutions have the same objective value w.r.t. the resource sharing problem. As indicated in \cref{subfig:only_local_congestion}, however, the maximum congestion might be caused by local effects that are beyond reach for global optimization. A practicable algorithm must be resilient to such local hotspots and produce solutions that are close to optimal on the set of remaining resources. The solution in \cref{subfig:only_local_congestion} is clearly to be preferred over that in \cref{subfig:completely_congested}. In practice, the ideal outcome might be a solution $x \in \mathcal{X}$ that minimizes the maximal entry, and among such solutions, it minimizes the second-highest entry, and among those, it minimizes the third-highest, and so on. This concept is known as \emph{decreasingly minimal} \cite{frank_murota_part1}.
\begin{definition}
Let $x \in \mathbb{R}^m$. We denote by $\sort{x}$ the vector $x$ with entries sorted in decreasing order. We introduce the decreasing preorder, by defining for $x,y \in \mathbb{R}^m$, $x \decleq y$ if either $\sort{x} = \sort{y}$ or there exists $k \in \{1,...,m\}$ such that $(\sort{x})_k < (\sort{y})_k$ and $(\sort{x})_j = (\sort{y})_j$ for all $j < k$. \par
Given a set $X \subseteq  \mathbb{R}^m$, we say that an element $x \in X$ is \emph{decreasingly minimal} if $x \decleq y$ holds for all $y \in X$.
\end{definition}
It is well-known that decreasingly minimal elements exist and that they are unique in non-empty closed convex subsets of $\mathbb{R}^m_{\geq 0}$ \cite{lexicographic_properties}. Therefore, the following problem is well-defined.
\begin{problem}
Given an instance $\mathcal{X}$ of the resource sharing problem, find the decreasingly minimal element $\lambda \in \mathcal{X}$ and a decomposition $\lambda = \sum_{C \in \mathcal{C}} x^{(C)}$ with $x^{(C)} \in X_C$ for all $C \in \mathcal{C}$.
\end{problem}

\subsection{Our contribution}
In this work, we present a fully polynomial-time approximation scheme for the primal as well as the dual of the resource sharing problem. Our algorithm is an extension of the algorithm by Müller, Radke, and Vygen \cite{min_max_resource_sharing_mrv}. We replace their binary search with a successive scaling/restarting approach. By bounding the running time with a geometric series, we can reduce their number of oracle calls, which was $\mathcal{O}((n+m)\log m(\log \log m+\delta^{-2}))$, to $\mathcal{O}(\frac{n+m}{\delta^2} \log m)$. This improves on the best-known number of oracle calls even in the special case of the maximum concurrent flow problem.\par
The core algorithm runs in $T$ \emph{phases}. In each phase $t=1,...,T$, a solution $s_C^{(t)}$ for every customer $C \in \mathcal{C}$ is computed. This is done by querying the oracle function with a price vector $y \in \mathbb{R}^m_{\geq 0}$. However, a solution $b:=f_C(y)$ is accepted by at most a fraction of $1/\norm{b}_{\infty}$, following the width-reduction technique of Garg and Könemann \cite{garg_koenemann_mcf}. At the end, a convex combination of the $s_C^{(1)},...,s_C^{(T)}$ is returned. This general approach is described as \cref{alg:generalized_resource_sharing_core}. We prove, using a result by Klein and Young \cite{klein_young_lower_bound_iterations}, that any algorithm that follows this general outline requires $\Omega(\frac{n+m}{\delta^2}\log m)$ oracle calls to compute a $(1+\delta)$-approximate solution (for a range of parameters as described in \cref{theo:lower_bound_iterations_rs}). As this is independent of the choice of the prices, this proves that --- in a certain sense --- multiplicative price updates are optimal, and that no warm-start analysis (i.e. reducing the running time by starting with a close to optimal dual solution) of the algorithm is possible.\par
Further, we introduce the concept of \emph{local weak duality} (\cref{def:local_weak_duality}). If this is satisfied, we can prove stronger bounds on $\max_{r \in S} x_r$ for our computed solution $x \in \mathcal{X}$ and certain subsets $S \subseteq \mathcal{R}$. That provides a theoretical counterpart to the empirical observation that the algorithm optimizes (close to) independent parts of the instance separately. Moreover, it allows concluding that --- with an exact block solver --- our algorithm always computes a solution that is approximately decreasingly minimal on the two highest entries. We also construct an example that proves that this statement is not true for the three highest entries in general.

\subsection{Related work}
The resource sharing problem originates from the maximum concurrent flow problem. For this special case, Shahrokhi and Matula \cite{shahrokhi_matula} introduced the idea to use Dantzig-Wolfe-type algorithms with an exponential weight function. Another important special case of this problem is \emph{fractional packing}, which can be described as follows. Given a polyhedron $P \subseteq \mathbb{R}^k$, a matrix $A \in \mathbb{R}^{m \times k}$ that satisfies $Ax \geq 0$ for all $x \in P$, and a vector $b \in \mathbb{R}^m_{> 0}$, find an $x \in P$ that satisfies $Ax \leq b$. The width of the instance is defined as $\rho := \max_{x \in P} \max_{i=1,...,m} (Ax)_i/b_i$. Plotkin, Shmoys, and Tardos \cite{plotkin_tardos_framework} studied this problem in the context of Dantzig-Wolfe-type algorithms. Their results were generalized and improved multiple times \cite{fleischer,garg_koenemann_mcf,grigoriadis_many_blocks,jansen_zhang,Khandekar_phd,young_nearly_linear_time,min_max_resource_sharing_mrv}. An important idea in this line of work is a step-size technique introduced by Garg and Könemann \cite{garg_koenemann_mcf} and extended by Fleischer \cite{fleischer}, which is used to design algorithms with width-independent running times. \par 
The general version of the block-angular min-max resource sharing problem, as it is the subject of this work, was studied first by Grigoriadis and Khachiyan \cite{grigoriadis_many_blocks}. In their formulation, one is given non-empty closed convex blocks $B_C$ for each $C \in \mathcal{C}$, and resource allocation functions $g^{(C)}:B_C \rightarrow \mathbb{R}^m_{\geq 0}$, which are convex and non-negative in each coordinate. Then, one seeks to compute $\min \{\max_{r \in \mathcal{R}} \sum_{C \in \mathcal{C}} g^{(C)}(b_C)_r : b_C \in B_C \, \forall C \in \mathcal{C}\}$. Their optimization oracle solves $\min_{b_C \in B_C} \inp{y}{g^{(C)}(b_C)}$ for a given price vector $y \in \mathbb{R}^m_{\geq 0}$. It is easy to see that this formulation fits into our framework by defining $X_C$ as the convex hull of $g^{(C)}(B_C)$. The currently fastest known algorithm for the general case is due to Müller, Radke, and Vygen \cite{min_max_resource_sharing_mrv}. They present a sequential algorithm that computes a solution of value at most $\sigma(1+\delta)$ within $\mathcal{O}((n+m) \log m (\delta^{-2} + \log \log m))$ oracle calls.  All of the mentioned approaches can be interpreted as variants of the multiplicative weights update method \cite{arora_multiplicative_weights}.\par
Klein and Young \cite{klein_young_lower_bound_iterations} studied lower bounds on the number of iterations that are required by any Dantzig-Wolfe-type algorithm to compute a $(1+\delta)$-approximate solution for fractional packing. They provide an asymptotic width-dependent bound of $\Omega\left(\min \left\{\rho \frac{\log m}{\delta^2}, m^{1/2-\gamma}\right\}\right)$ (for any fixed $\gamma \in (0,1/2)$). This matches known upper bounds precisely for a range of parameters. If the width of the instance is unbounded, it is easy to see that $\Omega(m)$ oracle calls are required to compute a constant-factor approximation \cite{grigoriadis_khachiyan_ppdd}. If one is not restricted to algorithms of the Dantzig-Wolfe-type, then it is known how to avoid the $\delta^{-2}$ dependence on the running time for the case of fractional packing \cite{zhu_orecchia_packing_covering,bienstock_epsilon_bound}.\par 
Decreasingly minimal solutions to optimization problems appear under many different names in the literature, such as lexicographically optimal \cite{fujishige_base_polyhedron,meggido_flows}, egalitarian \cite{dutta_ray_egalitarian_allocations}, fair \cite{frank_murota_fair_flows,lexicographic_properties} and more recently in a line of work by Frank and Murota --- who study the integral case --- as decreasingly minimal \cite{frank_murota_part1,frank_murota_part2}. We are going to use their notation in the following. It is known how to find such solutions with linear programming techniques \cite{orlin_lexmin_lp_approach,lexicographic_properties}. In our case, since $\mathcal{X}$ is a non-empty closed convex subset of $\mathbb{R}^m_{\geq 0}$, it contains a unique decreasingly minimal element \cite{lexicographic_properties}. A concept related to decreasing minimization is that of \emph{majorization} \cite{hardy_littlewood_polya}. If the set of feasible solutions contains a least majorized element, it is also the decreasingly minimal element and can be extracted as the minimum of non-decreasing separable convex functions \cite{tamir_least_majorized}. We are not aware of results w.r.t. decreasingly minimal solutions in the context of Dantzig-Wolfe-type algorithms.

\subsection{Outline}
\begin{algorithm}[t]
  \caption{Standard block-coordinate descent with restricted steps}
  \begin{algorithmic}[1]
      \For{$t = 1,..,T$}
      \For{$C \in \mathcal{C}$}
      \State $\alpha \gets 0$, $s_C^{(t)} \gets 0$
      \While{$\alpha < 1$}
      \State Choose $y \in \mathbb{R}^m_{\geq 0}$
      \State $b \gets f_C(y)$ \label{line:block_solver_call_rs} 
      \State $\xi \gets \min \{1 - \alpha, 1/\norm{b}_{\infty} \}$ \label{line:restricted_oracle_call_rs}
      \State $s_C^{(t)} \gets s_C^{(t)}+\xi b$ 
      \State $\alpha \gets \alpha + \xi$
      \EndWhile
      \EndFor
      \State $s^{(t)} \gets \sum_{C \in \mathcal{C}} s_C^{(t)}$ \label{line:combine_solutions}
      \EndFor
      \State \Return a convex combination of $s^{(1)},...,s^{(T)}$ 
  \end{algorithmic}
  \label{alg:generalized_resource_sharing_core}
\end{algorithm}

As mentioned before, \cref{alg:generalized_resource_sharing_core} describes a general class of Dantzig-Wolfe-type algorithms, which work in $T \in \mathbb{N}$ phases and process each customer in every phase following the restricted step-size rule due to Garg and Könemann (\cref{line:restricted_oracle_call_rs}). The bound $1/\norm{b}_{\infty}$ comes from the assumption that $\lambda^*$ is "close to" 1. Usually, the instance $\mathcal{X}$ has to be scaled, before a version of \cref{alg:generalized_resource_sharing_core} is run. \par
The algorithm by Müller, Radke, and Vygen is a special case of this algorithm that works with multiplicative price updates. It starts with a uniform price vector $y= \mathbbm{1} \in \mathbb{R}^m$ and, after every oracle call, it updates the prices $y_r \gets y_r \exp(\epsilon \xi b_r)$, where $\epsilon > 0$ is a fixed parameter of the algorithm. At the end, the average solution $x^{(T)} := \frac{1}{T}\sum_{t=1}^T s^{(t)}$ is returned. We will refer to this algorithm as the \emph{core algorithm} in the following.\\
The standard tool to prove primal convergence is weak duality: For $y \in \mathbb{R}^m_{\geq 0}$, it holds that $ \sum_{C \in \mathcal{C}} \inp{y}{f_C(y)}\leq \sigma \min_{x \in \mathcal{X}} \inp{y}{x} \leq \sigma \norm{y}_1 \min_{x \in  \mathcal{X}} \norm{x}_{\infty} = \sigma \norm{y}_1 \lambda^*$. This is sufficient to bound $\max_{r \in \mathcal{R}} x_r^{(T)}$. However, we aim to prove stronger bounds on $\max_{r \in S}x_r^{(T)}$ for certain subsets of resources $S \subseteq \mathcal{R}$. To this end, we introduce the concept of \emph{local weak duality}, which generalizes weak duality. We briefly describe the intuition behind this notion. Given $y \in \mathbb{R}^m_{\geq 0}$, one may consider the \emph{local} objective value $\sum_{C \in \mathcal{C}} \sum_{r \in S} y_r f_C(y)_r$ of the oracles on $S$ (e.g. the cost of the Steiner trees restricted to a subset of edges in the global routing case). A local analog to weak duality is given if this objective value can be bounded by  $\mu \sum_{r \in S}y_r$ for some $\mu > 0$, independently of $y_r$ for $r  \in \mathcal{R}\setminus S$ (the prices on the remaining edges). The following definition includes a different price vector for every customer, which is necessary to deal with the sequential price updates of the core algorithm. Then, the upper bound on the local objective value is defined using the point-wise maximum of these prices.
\begin{definition}
\label{def:local_weak_duality}
We say that an instance $\mathcal{X}$ of the resource sharing problem satisfies \emph{local weak duality w.r.t. a subset of resources $S \subseteq \mathcal{R}$ and $\mu \geq 0$} if for any collection of non-negative price vectors $(y^{(C)})_{C \in \mathcal{C}} \subseteq \mathbb{R}^m_{\geq 0}$  it holds that 
\begin{equation} \sum_{C \in \mathcal{C}} \sum_{r \in S} y^{(C)}_r f_C(y^{(C)})_r   \leq \mu \sum_{r \in S} \max_{C \in \mathcal{C}}y^{(C)}_r. \label{eq:monotone_weak_duality}\end{equation}
\end{definition}
This definition generalizes weak duality, in the sense that every instance satisfies local weak duality w.r.t. $S = \mathcal{R}$ and $\mu = \sigma \lambda^*$. \par
In \cref{subsec:analysis_normalized_instances}, we analyze the core algorithm on normalized instances (instances on which $\lambda^*$ is known up to a constant factor). We prove primal-dual convergence and that, under local weak duality, $\max_{r \in S} x^{(T)}_r$ is close to $\mu$. Using a fast restarting/scaling algorithm, which computes a constant-factor approximation in $\mathcal{O}((n+m) \log m)$ oracle calls as described in \cref{subsec:constant_factor_approximation}, we can transfer these results to the general setting without increasing the asymptotic number of oracle calls. This is summarized in the following main theorem.

\begin{theorem}[Main Theorem]
\label{theo:rs_main_theorem}
Let $\mathcal{X}$ be an instance of the resource sharing problem and $\delta \in (0,1]$. One can compute a primal solution $x \in \mathcal{X}$ and a dual solution $z \in \Delta_m$ satisfying
\[ \norm{x}_{\infty} \leq (1+\delta) \sigma \lambda^*, \qquad\qquad \min_{x \in \mathcal{X}} \inp{z}{x} \geq (1-\delta) \frac{\lambda^*}{\sigma},\]
\[ \text{and} \qquad \max_{r \in S} x_r \leq \mu + \delta \max\{\lambda^*, \mu \} \qquad \forall S,\mu \text{ s.t. \eqref{eq:monotone_weak_duality} holds},\]
using  $\mathcal{O}\left(\frac{(n+m)}{\delta^2} \log m \right)$ oracle calls, and further operations taking polynomial time.
\end{theorem}
In \cref{sec:dec_min_two}, we show that this result implies that the computed solution is close to decreasingly minimal on the two highest entries in the following sense.
\begin{corollary}
\label{cor:dec_min_two}
Let $\lambda \in \mathcal{X}$ be decreasingly minimal. If the block solver is exact, i.e. $\sigma = 1$, then the returned solution $x\in \mathcal{X}$ satisfies
\[ (\sort{x})_1\leq (\sort{\lambda})_1 +\delta \lambda^* \qquad \text{and} \qquad (\sort{x})_2 \leq (\sort{\lambda})_2 + \delta \lambda^*.\]
\end{corollary}
This analysis is best possible for the general case. \cref{theo:rs_all_entries_large} shows that an analogous version of \cref{cor:dec_min_two} does not hold for the three highest entries.\par
Moreover, in \cref{sec:lower_bound_iterations}, we prove that our algorithm is optimal in the sense that any version of \cref{alg:generalized_resource_sharing_core} requires $\Omega(\frac{n+m}{\delta^2})$ oracle calls to construct a $(1+\delta)$-approximate solution, even if $\lambda^*=1$ is known. 

\begin{theorem}
\label{theo:lower_bound_iterations_rs}
For every $\gamma \in (0,1/2)$ there exist constants $K_{\gamma}, \tau_{\gamma} > 0$, such that for every $m > K_{\gamma}$, $n \in \mathbb{N}$, there exists an instance of the resource sharing problem with $n+2$ customers, $2m$ resources and $\lambda^* = 1$, such that for any $\delta \in (0,1/10)$, any version of \cref{alg:generalized_resource_sharing_core} requires
\[ \tau_{\gamma} (n+m) \min\left\{\frac{\log m}{\delta^2}, m^{1/2 - \gamma} \right\} \] 
oracle calls to compute a $(1+\delta)$-approximate solution.
\end{theorem}

\subsection{Remarks on notation}
\label{subsec:remarks_notation}
We briefly point out how the original formulation of Grigoriadis and Khachiyan \cite{grigoriadis_many_blocks} (also used in \cite{jansen_zhang,min_max_resource_sharing_mrv}) fits into our framework. They consider general closed non-empty convex sets $B_C$ for every customer $C \in \mathcal{C}$. For each resource $r \in \mathcal{R}$ and each customer a convex non-negative resource consumption function $g^{(C)}_r : B_C \rightarrow \mathbb{R}_{\geq 0}$ is introduced.  We write $g^{(C)}(x) := (g^{(C)}_r(x))_{r \in \mathcal{R}}$. One seeks to find 
\begin{equation} \lambda^* := \min \left\{ \max_{r \in \mathcal{R}} \sum_{C \in \mathcal{C}} g_r^{(C)}(x_C) : x_C \in B_C \right\}.  \label{eq:standard_resource_sharing_def}\end{equation}
This formulation emphasizes that we usually work with a feasible region $B_C$ (e.g. the Steiner tree polytope in global routing) which specifies the solutions for customer $C$. When a multitude of different resources is considered, the feasible resource allocations are not described by $B_C$. For this purpose, the resource consumption functions $g^{(C)}$ are introduced. They map a solution from $B_C$ to its corresponding resource usage.\\
In this setting, for a given $y \in \mathbb{R}_{\geq 0}$ the oracle functions return an element $f^{(C)}(y) \in B_C$ that satisfies $\inp{y}{g^{(C)}(f^{(C)}(y))} \leq \sigma \min_{x_C \in B_C} \inp{y}{g^{(C)}(x_C)}$. To adapt this to our setting, we may assume that the oracle functions return $g^{(C)}(f^{(C)}(y))$ and would like to set $X_C := g^{(C)}(B_C)$ as by \cref{eq:standard_resource_sharing_def}
\[ \lambda^* = \min_{x \in \mathcal{G}} \norm{x}_{\infty}, \]
where $\mathcal{G} := \sum_{C \in \mathcal{C}} g^{(C)}(B_C)$. However, $g^{(C)}(B_C)$ is not necessarily a convex set. We circumvent this obstacle by setting $X_C := conv(g^{(C)}(B_C))$. The following lemma implies that we still have an approximate block solver over $X_C$ and that the optimum objective value is unchanged.
\begin{lemma}
With the definitions above, for every $y \in \mathbb{R}^m_{\geq 0}$:
\begin{equation}
\min_{x \in g^{(C)}(B_C)} \inp{y}{x} = \min_{x \in X_C} \inp{y}{x}
    \label{eq:same_objective_value_on_convex_hull}
\end{equation}
and further for $\mathcal{X} := \sum_{C \in \mathcal{C}} X_C$ and $\mathcal{G}:= \sum_{C \in \mathcal{C}} g^{(C)}(B_C)$ we have 
\begin{equation}
\min_{x \in \mathcal{X}} \norm{x}_{\infty} = \min_{g \in \mathcal{G}} \norm{g}_{\infty}.
    \label{eq:same_min_on_convex_hull}
\end{equation}
\end{lemma}
\begin{proof}
\cref{eq:same_objective_value_on_convex_hull} holds because linear functions on convex sets are minimized on minimal faces. For \cref{eq:same_min_on_convex_hull} it is clear that "$\leq$" holds. On the other hand, let $x \in \mathcal{X}$ be such that the left-hand side of \cref{eq:same_min_on_convex_hull} is attained by $x$. We have $x = \sum_{C \in \mathcal{C}}x^{(C)}$ for some $x^{(C)} \in X_C$, which can be written as convex combinations 
\[ x^{(C)} = \sum_{k=1}^{K_C} \alpha_{C,k} g^{(C)}(b^{(C,k)})\]
for $b^{(C,k)} \in B_C$ and some $K_C \in \mathbb{N}$. Now by Jensen`s inequality we have for $b^{(C)} := \sum_{k=1}^{K_C} \alpha_{C,k} b^{(C,k)} \in B_C$ that 
\[ x^{(C)}_r = \sum_{k=1}^{K_C} \alpha_{C,k} g_r^{(C)}(b^{(C,k)}) \geq g_r^{(C)}(b^{(C)})\]
for every $r \in \mathcal{R}$. Therefore $\norm{x}_{\infty} \geq \norm{g}_{\infty}$ where $g:= \sum_{C \in \mathcal{C}} g^{(C)}(b^{(C)}) \in \mathcal{G}$. This shows "$\geq$" in \cref{eq:same_min_on_convex_hull}.
\end{proof}
Therefore, the definition in \cite{min_max_resource_sharing_mrv} is compatible with our formulation. We can work on $\mathcal{X}$ instead of $\mathcal{G}$, as we have an approximate oracle for linear optimization over $\mathcal{X}$ by \cref{eq:same_objective_value_on_convex_hull}, and the optimum objective value of the resource sharing problem is unchanged by \cref{eq:same_min_on_convex_hull}. Once we have a solution in $\mathcal{X}$ we can easily reconstruct one in $\mathcal{G}$ which has at most the same objective value as in the proof above.

\section{Proof of the main theorem}
\label{sec:resource_sharing_core_algorithm}
\subsection{Analysis on normalized instances}
\label{subsec:analysis_normalized_instances}
First, we prove the main theorem for normalized instances, meaning on those for which $\lambda^* \in [c,1]$ is known for a constant $c >0$. Later, in \cref{subsec:constant_factor_approximation} we show how to remove this assumption.

\begin{algorithm}[t]
  \caption{Core Resource Sharing Algorithm with parameters $\epsilon, T$}
  \begin{algorithmic}[1]
  	 \State $y \gets \mathbbm{1} \in \mathbb{R}^m$ 
      \For{$t = 1,..,T$}
      \For{$C \in \mathcal{C}$}
      \State $\alpha \gets 0$
      \State $s_C^{(t)} \gets 0$
      \While{$\alpha < 1$}
      \State $b \gets f_C(y)$ \label{line:block_solver_call_rs}
      \State $\xi \gets \min \{1 - \alpha, 1/\norm{b}_{\infty} \}$ \label{line:restricted_oracle_call_rs}
      \State $y_r \gets y_r \exp(\epsilon \xi b_r)$ $\, \forall r \in \mathcal{R}$ \label{line:price_vector_sequential_update}
      \State $s_C^{(t)} \gets s_C^{(t)}+\xi b$ 
      \State $\alpha \gets \alpha + \xi$
      \EndWhile
      \EndFor
      \State $s^{(t)} \gets \sum_{C \in \mathcal{C}} s_C^{(t)}$ \label{line:combine_solutions}
      \EndFor
      \State \Return $\frac{1}{T} \sum_{t=1}^T s^{(t)}$
  \end{algorithmic}
  \label{alg:mrv_resource_sharing_core}
\end{algorithm}
\cref{alg:mrv_resource_sharing_core} describes the core algorithm that is due to Müller, Radke, and Vygen \cite{min_max_resource_sharing_mrv}. As mentioned earlier, it is a version of \cref{alg:generalized_resource_sharing_core}. The algorithm runs in $T$ \emph{phases} (iterations of the outer for-loop). In each phase $t=1,...,T$, a solution $s_C^{(t)} \in X_C$ for each customer $C \in \mathcal{C}$ is constructed. These solutions are combined to a solution $s^{(t)} \in \mathcal{X}$ (\cref{line:combine_solutions}). At the end, the average solution across all phases $\frac{1}{T} \sum_{t=1}^T s^{(t)} \in \mathcal{X}$ is returned.\par
The solutions $s_C^{(t)}$ are constructed in the inner for-loop. When customer $C \in \mathcal{C}$ is processed, the value $\alpha$ indicates the fraction of a solution that was collected in the current phase. At the end of the while-loop, we have $\alpha = 1$. As long as $\alpha < 1$, we ask the oracle function of customer $C$ to return a solution for the current price vector $y$ (\cref{line:block_solver_call_rs}). In the next step, \cref{line:restricted_oracle_call_rs}, we decide the coefficient $\xi$ with which $b$ should be added to the solution $s_C^{(t)}$ which is then added to $\alpha$. It is bounded by $1-\alpha$ (as we want $\alpha =1$ at termination of the while-loop) and $\frac{1}{\norm{b}_{\infty}}$. The interpretation of the second bound is that we are more "careful" with solutions $b$ that have a high entry $b_r$ for some $r \in \mathcal{R}$ and add them to the current solution only with a small fraction. The algorithm hides the implicit estimate $\lambda^* = 1$. Fast convergence with the core algorithm is only achieved when the instance is scaled such that $\lambda^*$ is "close" to 1. From a technical perspective, the reason to employ the bound $\xi \leq \frac{1}{\norm{b}_{\infty}}$ is that $\xi b_r \leq 1$ always holds. This allows to bound the price increase in \cref{line:price_vector_sequential_update} by a linear upper bound
\begin{equation}
    y_r \exp( \epsilon \xi b_r) \leq y_r \left( 1+ \frac{e^{\epsilon} -1}{\epsilon} \epsilon \xi b_r \right) = y_r + (e^{\epsilon} -1) \xi b_r, \label{eq:linear_exponential_function}
\end{equation}
which holds by convexity of the exponential function and the fact that $\epsilon \xi b_r \leq \epsilon$ is guaranteed. Inequality \eqref{eq:linear_exponential_function} is crucial to prove the convergence of the algorithm.
\begin{remark} In \cite{min_max_resource_sharing_mrv} it is not required to construct a complete solution for a customer before considering a different one in the same phase. All results presented here hold also for their version of the algorithm. \end{remark} 

We fix the following notation for the analysis of the algorithm:
\begin{itemize}
    \item[(i)] We write $x^{(t)} := \frac{1}{t} \sum_{p=1}^t s^{(p)} \in \mathcal{X}$ for the average solution after phase $t$ is completed.
    \item[(ii)] We denote with $y^{(t)}$ the price vector $y$ after phase $t$ is completed.
\end{itemize}
In principle, it would be possible that a phase does not terminate because of the while-loop. It will become evident with \cref{lem:bound_number_of_oracle_calls_rs} and in the proof of \cref{lem:rs_bound_price_increase} that this does not happen and thus the values $x^{(t)}$ and $y^{(t)}$ are well-defined. 
Note that $y_r^{(t)} = \exp(\epsilon \sum_{p=1}^t s_r^{(p)}) = \exp(\epsilon t x_r^{(t)})$ holds for all $r \in \mathcal{R}$ and $t =1,...,T$.  Again, the normalized prices are dual solutions to the resource sharing problem. However, we do not normalize them in this algorithm, because it is convenient for the analysis to have prices that are monotone increasing. Normalization does not influence the behavior of the algorithm in any way. We can rewrite the relation between primal and dual solution equivalently as
\begin{equation}
x_r^{(t)} = \frac{1}{\epsilon t} \log y_r^{(t)}.
    \label{eq:price_solution_transform}
\end{equation}
This can be used to deduce a simple bound for any $S \subseteq \mathcal{R}$:
\begin{equation}
\max_{r \in S} x^{(t)}_r = \max_{r \in S} \frac{1}{\epsilon t} \log y_r^{(t)} \leq \frac{1}{\epsilon t} \log \sum_{r \in S} y^{(t)}_r.
    \label{eq:bound_max_usage_by_price}
\end{equation}
Especially, $\norm{x^{(t)}}_{\infty} \leq \frac{1}{\epsilon t} \log \norm{y^{(t)}}_1$ holds. We denote the dual objective values that correspond to the price vectors $y^{(t)}$ by 
\begin{equation} \Theta_t := \min_{x \in \mathcal{X}} \frac{\inp{y^{(t)}}{x}}{\norm{y^{(t)}}_1} = \frac{\sum_{C \in \mathcal{C}} opt_C(y^{(t)})}{\norm{y^{(t)}}_1}.
        \label{eq:dual_values}
\end{equation}
Because of the while-loop, it is not immediately clear how to bound the number of oracle calls of \cref{alg:mrv_resource_sharing_core}. We provide a bound that relates the number of oracle calls to the price vector $y^{(t)}$. This improves on the bound in \cite{min_max_resource_sharing_mrv}, which was $tn + \frac{m}{\epsilon} \log \norm{y^{(t)}}_1$.
\begin{lemma}
\label{lem:bound_number_of_oracle_calls_rs}
The number of oracle calls of \cref{alg:mrv_resource_sharing_core} until termination of phase $t =1,...,T$ is bounded by 
\[ tn+ \frac{m}{\epsilon} \log \frac{\norm{y^{(t)}}_1}{m}.\]
\end{lemma}
\begin{proof}
It is clear that there are at most $tn$ oracle calls for which $\xi$ is chosen as $\xi = 1 - \alpha$. It is more difficult to bound the number of oracle calls for which $\xi$ is chosen as $\xi = \frac{1}{\norm{b}_{\infty}}$. In this case, we know that there is a resource $r \in  \mathcal{R}$, such that $\xi b_r = 1$, and the price update in \cref{line:price_vector_sequential_update} reads $y_r = y_r \exp(\epsilon)$. For $r \in \mathcal{R}$ let $\kappa^{(t)}_r$ denote the number of oracle calls for which this occurs before termination of phase $t$. We get the bound $y^{(t)}_r \geq \exp(\epsilon \kappa^{(t)}_r)$. The total number of oracle calls with $\xi = \frac{1}{\norm{b}_{\infty}}$ is bounded by $\sum_{r \in \mathcal{R}} \kappa^{(t)}_r$. We can bound this sum in the following way:
\begin{align*}
    \sum_{r \in \mathcal{R}} \kappa^{(t)}_r & \leq \frac{1}{\epsilon} \sum_{r \in \mathcal{R}} \log y^{(t)}_r 
      = \frac{1}{\epsilon} \log \prod_{r \in \mathcal{R}} y^{(t)}_r 
      = \frac{m}{\epsilon} \log \left(\prod_{r \in \mathcal{R}} y^{(t)}_r\right)^{(1/m)} 
      \leq \frac{m}{\epsilon} \log \frac{\norm{y^{(t)}}_1}{m},
     \end{align*}
     where we used the inequality between geometric and arithmetic mean in the last step.
\end{proof}

\cref{eq:bound_max_usage_by_price} and \cref{lem:bound_number_of_oracle_calls_rs} suggest that it is enough to find a good bound on $\norm{y^{(T)}}_1$ to bound the maximum entry of $x^{(T)}$ as well as the total number of oracle calls. Further, if we can derive a bound on $\sum_{r \in S} y_r^{(t)}$ for $S \subseteq \mathcal{R}$ we can derive a bound on $\max_{r \in S} x_r^{(t)}$. This leads to the most important statement of the analysis,  \cref{lem:rs_bound_price_increase}. 
\begin{lemma}[Upper bound on the price increase]
\label{lem:rs_bound_price_increase}
Let $\mathcal{X}$ be an instance of the resource sharing problem that satisfies local weak duality w.r.t.  $S \subseteq \mathcal{R}$ and $\mu > 0$. Let $\eta := \exp(\epsilon) -1$. Assume that $\eta \mu < 1$. Then for all $t =0,1,...,T$, it holds that
\[ \sum_{r \in S} y_r^{(t)}  \leq |S| \exp\left( \frac{t\eta \mu}{1- \eta \mu} \right).\]
\end{lemma}
\begin{proof}
We prove the bound by induction on $t$. For $t=0$, we have $y^{(0)} = \mathbbm{1}$ and thus $\sum_{r \in S} y_1^{(0)} = |S|$ holds. Now let us consider a phase $t > 0$. The goal is now to bound $\sum_{r \in S} y^{(t)}_r/ \sum_{r \in S} y^{(t-1)}_r \leq \exp( \frac{\eta \mu}{1-\eta \mu}) $, which proves the desired bound. \\
To do so, we first bound the price increase for single oracle calls. Consider a number $\kappa_t \in \mathbb{N}$ of oracle calls in phase $t$. For $j =0,....,\kappa_t$ we fix the following notation: $b^{(t,j)}, \xi^{(t,j)}, C(t,j)$ are the values of $b$, $\xi$, respectively $C$ of the $j$-th oracle call in phase $t$; $y^{(t,j)}$ denotes the price vector $y$ after the update in \cref{line:price_vector_sequential_update} following the $j$-th oracle call. Using Inequality \eqref{eq:linear_exponential_function}, we can bound for each $r \in \mathcal{R}:$
\[ y_r^{(t,j)} = y_r^{(t,j-1)} \exp(\epsilon \xi^{(t,j)} b_r^{(t,j)}) \leq y_r^{(t,j-1)} \left( 1 + \eta \xi^{(t,j)} b_r^{(t,j)}\right).\]
Summing over all $r \in S$, we  get
\begin{equation}
\sum_{r \in S} y_r^{(t,j)} \leq \sum_{r \in S}y_r^{(t,j-1)} + \eta \xi^{(t,j)} \sum_{r \in S} y_r^{(t,j-1)}b_r^{(t,j)}.
    \label{eq:single_iteration_price_increase_bound}
\end{equation}
Recall that $b^{(t,j)} = f_{C(t,j)}(y^{(t,j-1)})$. The goal is now of course to apply local weak duality. To this end, we take a price vector $y^{(C)}$ for each customer that achieves the largest local objective value, i.e. 
\[y^{(C)} \in \argmax_{y^{(t,j-1)}: C(t,j) = C} \sum_{r \in S} y_r^{(t,j-1)} f_C(y^{(t,j-1)})_r. \]
Then we can estimate, since $\sum_{j: C(t,j)=C} \xi^{(t,j)} \leq 1$ for all $C \in \mathcal{C}$:
\begin{equation} \sum_{r \in S} y_r^{(t, \kappa_t)}  -\sum_{r \in S} y_r^{(t-1)}\leq  \sum_{j=1}^{\kappa_t}\eta \xi^{(t,j)} \sum_{r \in S} y_r^{(t,j-1)}b_r^{(t,j)} \leq \eta \sum_{C \in \mathcal{C}} \sum_{r \in S} y_r^{(C)} f_C(y^{(C)})_r. \label{eq:chain_of_inequalities} \end{equation}
Since the prices are increasing, local weak duality allows us to deduce
\[ \sum_{r \in S} y_r^{(t, \kappa_t)}  -\sum_{r \in S} y_r^{(t-1)} \leq  \eta \mu \sum_{r \in S} y_r^{(t,\kappa_t)}.\]
This shows that 
\[ \frac{\sum_{r \in S}y^{(t,\kappa_t)}_r}{\sum_{r \in S}y^{(t-1)}}\leq \frac{1}{1-\eta \mu} = 1+\frac{\eta \mu}{1-\eta \mu}\leq \exp \left(\frac{\eta \mu}{1-\eta \mu}\right),\]
 which holds for all $\kappa_t$. So, the sum of the prices in phase $t$ is uniformly bounded. Thus phase $t$ terminates and $y^{(t)}$ is well defined. The above inequality holds especially for the last oracle call in phase $t$, which proves the induction step.
\end{proof}

As pointed out earlier, every instance satisfies local weak duality w.r.t. $\mathcal{R}$ and $\sigma \lambda^*$. Therefore, the bound $\norm{y^{(t)}}_1\leq m \exp( t \eta \sigma \lambda^* / (1-\eta \sigma \lambda^*))$ that was stated in \cite{min_max_resource_sharing_mrv} follows. In this case, it is possible to insert the definition of the dual values $\sum_{C \in \mathcal{C}} opt_C(y^{(t)}) = \Theta_t \norm{y^{(t)}}_1$ to obtain a primal-dual bound.

\begin{lemma}
Assume that $\eta \sigma \lambda^* < 1$. Then for all $t=0,1,...,T$, it holds that
\[\norm{y^{(t)}}_1 \leq m \exp \left( \frac{\eta \sigma}{1-\eta \sigma \lambda^*} \sum_{p=1}^t \Theta_p \right) \]
\end{lemma}
\begin{proof}
This result can be shown analogously to the proof of the previous lemma. By inserting the definition of the dual values $\Theta_t$, one can derive a modification of \eqref{eq:chain_of_inequalities} (in the case $S = \mathcal{R}$):
\[ \norm{y^{(t)}}_1 - \norm{y^{(t-1)}}_1 \leq \eta \sum_{C \in \mathcal{C}} \sum_{r \in \mathcal{R}} y_r^{(C)} f_C(y^{(C)})_r \leq \eta \sigma \sum_{C \in \mathcal{C}} opt_C(y^{(t)}) = \eta \sigma \Theta_t \norm{y^{(t)}}_1.\]
It follows that (as $\Theta_t \leq \lambda^*$ by weak duality)
\[ \frac{\norm{y^{(t)}}_1}{\norm{y^{(t-1)}}_1} \leq \frac{1}{1- \eta \sigma \Theta_t} = 1+\frac{\eta \sigma \Theta_t}{1-\eta \sigma \Theta_t}\leq 1+\frac{\eta \sigma \Theta_t}{1-\eta \sigma \lambda^*}\leq \exp\left( \frac{\eta \sigma \Theta_t}{1-\eta \sigma \lambda^*}\right).\]
The claim follows by inductive application of the inequality above.
\end{proof}
\begin{remark}
Assuming $\eta \sigma \lambda^* < 1$, \cref{lem:rs_bound_price_increase} yields that the number of oracle calls up to phase $t =1,...,T$ is bounded by  $t (n+ c m)$, where $c$ is a constant that depends on $\epsilon, \sigma$ and $\lambda^*$.
\end{remark}
\cref{lem:rs_bound_price_increase} can be used to bound the primal as well as the dual error easily.
\begin{theorem}[Bound on the primal and dual error]
\label{theo:primal_dual_error}
Assume $\epsilon \in (0,1]$, $\eta \sigma \lambda^* < 1$. For every $t = 1,...,T$ the primal solution $x^{(t)}$ satisfies
\begin{equation}
\norm{x^{(t)}}_{\infty} \leq \frac{\log m}{\epsilon t} +  \frac{1+\epsilon}{1-\eta \sigma \lambda^*} \sigma \lambda^*.
    \label{eq:primal_error}
\end{equation}
If $\mathcal{X}$ satisfies local weak duality w.r.t. $S \subseteq \mathcal{R}$ and $\mu > 0$, then 
\begin{equation} \max_{r \in S} x^{(t)}_r \leq \frac{\log |S|}{\epsilon t} + \frac{1+\epsilon}{1-\eta \sigma \lambda^*}\mu. \label{eq:bound_local_weak_duality}\end{equation}
Moreover, the average dual solution $z^{(t)} := \frac{1}{t} \sum_{p=1}^t \frac{y^{(p)}}{\norm{y^{(p)}}_1}\in \Delta_m$ satisfies
\begin{equation}
    \min_{x \in \mathcal{X}} \inp{z^{(t)}}{x} \geq \frac{1- \eta \sigma \lambda^*}{\sigma (1+\epsilon)} \left( \lambda^* - \frac{\log m}{\epsilon t}\right).
    \label{eq:dual_error}
\end{equation}
\end{theorem}
\begin{proof}
We first prove Inequality \eqref{eq:bound_local_weak_duality}. The bound on the primal error \eqref{eq:primal_error} is an application of this inequality. We can assume w.l.o.g. that $\mu \leq \sigma \lambda^*$ (otherwise we can derive \eqref{eq:bound_local_weak_duality} from the fact that local weak duality is satisfied w.r.t. $\mathcal{R}$ and $\sigma \lambda^*$), such that $\eta \mu < 1$. Note that $\eta = \exp(\epsilon) -1 \leq \epsilon(1+\epsilon)$ for $\epsilon \in (0,1]$. By \cref{lem:rs_bound_price_increase}, we have
\[ \max_{r \in S} x^{(t)}_r \leq \frac{1}{\epsilon t} \log \sum_{r \in S} y_r^{(t)} \leq \frac{\log |S|}{\epsilon t} + \frac{t\eta \mu}{\epsilon t (1- \eta \mu)}\leq \frac{\log |S|}{\epsilon t} + \frac{1+\epsilon}{1-\eta \mu} \mu.\]
We can similarly bound the dual error. We know that $\lambda^* \leq \norm{x^{(t)}}_{\infty}$ must hold. Using again the \cref{eq:bound_max_usage_by_price}, we have $\lambda^* \leq \frac{1}{\epsilon t} \log \norm{y^{(t)}}_1$. With \cref{lem:rs_bound_price_increase} we can derive 
\[ \lambda^* \leq \frac{\log m}{\epsilon t} + \frac{\eta \sigma}{\epsilon t(1- \eta \sigma \lambda^*)} \sum_{p=1}^t \Theta_p \leq \frac{\log m}{\epsilon t} +\frac{\sigma (1+\epsilon)}{t(1-\eta \sigma \lambda^*)} \sum_{p=1}^t \Theta_p.\]
This can be used to obtain a lower bound on the average dual objective value of 
\[ \frac{1}{t} \sum_{p=1}^t \Theta_p  \geq \frac{1- \eta \sigma \lambda^*}{\sigma (1+\epsilon)} \left( \lambda^* - \frac{\log m}{\epsilon t}\right).\]
To conclude the argument and prove Inequality \eqref{eq:dual_error}, we observe that the average dual solution satisfies
\[ \min_{x \in \mathcal{X}} \inp{z^{(t)}}{x} = \min_{x \in \mathcal{X}} \inp{\frac{1}{t}\sum_{p=1}^t \frac{y^{(p)}}{\norm{y^{(p)}}_1}}{x} \geq \frac{1}{t}\sum_{p=1}^t \min_{x \in \mathcal{X}} \frac{\inp{y^{(p)}}{x}}{\norm{y^{(p)}}_1} = \frac{1}{t} \sum_{p=1}^t \Theta_p.\]
\end{proof}

The discussion above allows us to formulate the main theorem on normalized instances.
\begin{theorem}[Main theorem on normalized instances]
\label{theo:rs_normalized_main_theorem}
Assume that there is a constant $c > 0$, such that $\lambda^*  \in [c,1]$. Let $\delta \in (0,1]$. Then the core algorithm computes a primal solution $x^{(T)} \in  \mathcal{X}$ and a dual solution $z^{(T)} \in \Delta_m$ satisfying
\[ \norm{x^{(T)}}_{\infty} \leq (1+\delta) \sigma \lambda^*, \qquad \min_{x \in \mathcal{X}} \inp{z^{(T)}}{x} \geq (1-\delta) \frac{\lambda^*}{\sigma}\]
\[ \text{and} \qquad \max_{r \in S} x^{(T)}_r \leq \mu + \delta \max\{ \lambda^*, \mu\} \qquad \forall S,\mu \text{ s.t. local weak duality holds},\]
using  $\mathcal{O}\left(\frac{(n+m)}{\delta^2} \log m \right)$ oracle calls.
\end{theorem}
\begin{proof}
We choose $\epsilon = \frac{\delta}{8 \sigma}$ and $T = \left\lceil \frac{\log m }{2\sigma c \epsilon^2} \right\rceil$ and run the core algorithm. Note that (as $\epsilon \leq 1$) $\eta \leq 2 \epsilon$ and thus $\eta \sigma \lambda^* \leq \frac{\delta}{4} \lambda^* \leq \frac{\delta}{4}$. \cref{theo:primal_dual_error} guarantees that, if weak duality w.r.t. $S\subseteq \mathcal{R}$ and $\mu >0$ is satisfied (assuming again w.l.o.g. $\mu \leq \sigma \lambda^*$), 
\begin{flalign*}
    \max_{r \in S} x_r^{(T)} &  \leq \frac{\log m}{\epsilon T} + \frac{1+\epsilon}{1-\eta \mu} \mu \\
    & \leq 2\sigma c \epsilon + \frac{1+\epsilon}{1- \delta/4} \mu & \text{ Definition of $T$, bound on $\eta\mu \leq \eta \sigma \lambda^*$}\\
    & \leq \frac{\delta}{4}  \lambda^* + (1+\epsilon)\left(1+\frac{\delta}{2}\right) \mu & 1/(1-\delta/4) \leq 1+\delta/2, \text{ lower bound on $\lambda^*$}\\
    & \leq \frac{\delta}{4}  \lambda^* + \left(1+\frac{\delta}{8}\right) \left(1+\frac{\delta}{2}\right) \mu & \text{Definition of $\epsilon$}\\
    & \leq \frac{\delta}{4}  \lambda^* + \left(1+\frac{3}{4} \delta\right) \mu \\
    & \leq \mu + \delta \max\{\lambda^*, \mu\}.
\end{flalign*}
The bound on the primal error follows immediately. The dual value of the average dual solution $z^{(T)} := \frac{1}{T}\sum_{t=1}^T \frac{y^{(t)}}{\norm{y^{(t)}}_1}$ can also be bounded using \cref{theo:primal_dual_error} by 
\begin{flalign*}
    \min_{x \in \mathcal{X}} \inp{z^{(T)}}{x} & \geq \frac{1- \eta \sigma \lambda^*}{\sigma (1+\epsilon)}\left( \lambda^* - \frac{\log m}{\epsilon T}\right) \\
    & \geq \frac{1- \frac{\delta}{4}}{\sigma (1+\epsilon)} \left( \lambda^* - 2\sigma c \epsilon \right) & \text{ bound on $\eta \sigma \lambda^*$, Definition of $T$} \\
    & \geq \left(1- \frac{\delta}{4}\right) (1- \epsilon)\frac{\lambda^* - \frac{\delta}{4} c}{\sigma} & 1/(1+\epsilon) \geq 1-\epsilon, \text {Definition of $\epsilon$}\\
    & \geq \left(1-\frac{\delta}{4}\right)^2(1-\epsilon) \frac{\lambda^*}{\sigma} & \text{Definition of $\epsilon$, lower bound on $\lambda^*$}\\
    & \geq (1-\delta) \frac{\lambda^*}{\sigma}.
\end{flalign*}
These calculations prove the approximation guarantees. The number of oracle calls can be bounded, according to \cref{lem:bound_number_of_oracle_calls_rs}, by 
\begin{align*}
    nT + \frac{m}{\epsilon} \log \frac{\norm{y^{(T)}}_1}{m} \leq nT + \frac{m}{\epsilon} \frac{\eta \sigma T \lambda^*}{1- \eta \sigma \lambda^*} \leq T \left( n + m \frac{\sigma (1+\epsilon)}{1- \eta \sigma \lambda^*}\right)\leq  T \left( n + m \frac{\sigma (1 + \frac{1}{8})}{1-\frac{1}{4} }\right)  = T\left(n+\frac{3}{2}\sigma m\right),
\end{align*}
where we used $\lambda^* \leq 1$ and the bounds on $\epsilon$ and $\eta \sigma \lambda^*$.
Since $T \in \mathcal{O}(\frac{\log m}{\delta^2})$ the claimed bound on the number of oracle calls follows.
\end{proof}

\subsection{A constant-factor approximation using $\mathcal{O}((n+m) \log m)$ oracle calls}
\label{subsec:constant_factor_approximation}
We saw in the previous section, that the core algorithm (\cref{alg:mrv_resource_sharing_core}) can be used to compute a solution with objective value $\leq \sigma (1+\delta) \lambda^*$ using $\mathcal{O}(\log m\frac{n+m}{\delta^2})$ many oracle calls if one can guarantee that $\lambda^* \in [c,1]$ for some constant $c$. This can be done by computing a constant-factor approximation of $\lambda^*$ and scale the instance accordingly. In \cite{min_max_resource_sharing_mrv} this is achieved by a binary search/scaling approach within $\mathcal{O}((n+m) \log m \log \log m)$ oracle calls. 
The goal of this section is to provide a constant-factor approximation for the resource sharing problem within $\mathcal{O}((n+m) \log m)$ oracle calls. Combining our algorithm with the core algorithm proves the main theorem,  \cref{theo:rs_main_theorem}, in its general form. \par 
In the following (opposite to the previous assumption $\lambda^* \leq 1$), we assume that $\lambda^* \in [1, \sigma m]$. This can be guaranteed with $n$ oracle calls, by computing for each customer a solution to the uniform price vector $x := \sum_{C \in \mathcal{C}} f_C(\mathbbm{1})\in \mathcal{X}$. By definition of $\lambda^*$ and weak duality  we get the sandwich inequality 
\begin{equation} \lambda^* \leq \norm{x}_{\infty} \leq \norm{x}_1 \leq \sigma \norm{\mathbbm{1}}_1 \lambda^* = \sigma m \lambda^* .\label{eq:sandwich_lambda_l1_optimum}\end{equation}
We can then scale the instance to get a new one where the convex set $X_C$ is replaced by $\tilde{X_C} := \frac{\sigma m }{\norm{x}_{\infty}}X_C$ for each customer $C \in \mathcal{C}$. We also scale our solution $x$ to obtain $\tilde{x} := \frac{\sigma m }{\norm{x}_{\infty}} x$. Then we know that for the optimum in this scaled instance $\tilde{\lambda^*}$ it holds:
\[ \tilde{\lambda^*} \leq \norm{\tilde{x}}_{\infty} = \frac{\sigma m }{\norm{x}_{\infty}} \norm{x}_{\infty} = \sigma m.\]
Further, we have the lower bound
\[ \tilde{\lambda^*} = \frac{\sigma m}{\norm{x}_{\infty}} \lambda^* \underbrace{\geq}_{\text{\eqref{eq:sandwich_lambda_l1_optimum}}} \frac{\sigma m}{\norm{x}_{\infty}} \frac{\norm{x}_{\infty}}{\sigma m} = 1.\]
So $\tilde{\lambda^*}\in [1, \sigma m]$. We will avoid this notation in the following and assume that $\lambda^* \in [1, \sigma m]$. Note that scaling does not influence the approximation guarantees of the block solver. Next, we present the algorithm that achieves a constant-factor approximation. It is similar to the core algorithm, \cref{alg:mrv_resource_sharing_core}, but it works with an adaptive scaling approach, may discard the solution of some phases, and restart them. Our algorithm is given in pseudo-code as \cref{alg:fast_constant_factor_approximation}.
\begin{algorithm}[t]
  \caption{Constant-factor approximation for the resource sharing problem}
  \algnewcommand{\algorithmicgoto}{\textbf{go to}}%
\algnewcommand{\Goto}[1]{\algorithmicgoto~\ref{#1}}%
  \begin{algorithmic}[1]
  	 \State $y \gets \mathbbm{1} \in \mathbb{R}^m$  \Comment{price vector (at first uniform)}
  	 \State $\epsilon \gets \frac{1}{4\sigma}$ \Comment{Current epsilon for the price update}
  	 \State $\Lambda \gets 1$ \Comment{Current guess on $\lambda^*$}
      \For{$t = 1,.., T := \lceil \log m\rceil$} \Comment{Start phase}
      \State $y^{(t-1)} \gets y$ \Comment{Store the last price vector}
      \For{$C \in \mathcal{C}$} \label{begin_phase} \Comment{Process all customers}
      \State $\alpha \gets 0$ \Comment{"Fraction of solution that is collected for $C$"}
      \State $s_C^{(t)} \gets 0$ \Comment{Solution in this phase for customer $C$}
      \While{$\alpha < 1$}
      \State $b \gets f_C(y)$ \Comment{Call oracle for $C$ with price vector $y$}
      \State $\xi \gets \min \{1 - \alpha, \Lambda / \norm{b}_{\infty} \}$ \label{line:decide_xi_value} \Comment{Decide the "amount" with which $b$ is taken}
      \State $y_r \gets y_r \exp(\epsilon \xi b_r)$ $\, \forall r \in \mathcal{R}$\label{line:price_update_scaling} \Comment{multiplicative price update}
      \State $s_C^{(t)} \gets s_C^{(t)}+\xi b$  \Comment{Add $b$ to the current solution}
      \State $\alpha \gets \alpha + \xi$ \Comment{Update $\alpha$}
      \If{$\norm{y}_1 > m \exp \left(t\right)$} \label{line:check_price_bound}\Comment{Check price bound}
      \State $y \gets y^{(t-1)}$ \Comment{Reset the prices to those of the start of this phase}
      \State $\epsilon \gets \epsilon /2$ \Comment{Reduce $\epsilon$} \label{line:halve_epsilon}
      \State $ \Lambda \gets 2\Lambda$ \Comment{Increase the guess on $\lambda^*$}
      \State \Goto{begin_phase} \Comment{Restart phase, solutions from this phase are discarded}
      \EndIf
      \EndWhile
      \EndFor
      \State $s^{(t)} \gets \sum_{C \in \mathcal{C}} s^{(t)}_C$ \Comment{Store combined solutions from this phase}
      \EndFor
      \State
      \Return $\frac{1}{T}\sum_{t=1}^T s^{(t)}$ \Comment{Return the average solution across all phases}
  \end{algorithmic}
  \label{alg:fast_constant_factor_approximation}
\end{algorithm}

In \cref{line:check_price_bound} of \cref{alg:fast_constant_factor_approximation}, we have an if-statement that checks whether the sum of the prices exceeds a certain bound (that depends on the current phase $t$). If this bound is violated, all solutions from this phase are discarded, and phase $t$ is restarted, this time with a larger $\Lambda$ and a smaller $\epsilon$. The high-level idea is the following: We have an estimate on $\lambda^*$. This is given by $\Lambda$. If the bound in \cref{line:check_price_bound} is violated, we can conclude that our guess was too low and we double $\Lambda$. To compensate for this effect we halve $\epsilon$. The consequence is the following: We allow larger resource usage in the solutions returned by the oracle (\cref{line:decide_xi_value}) but increase the prices at a slower rate (\cref{line:price_update_scaling}).  At first, we note the following simple, but central observations.
At each step of \cref{alg:fast_constant_factor_approximation} (apart from \cref{line:halve_epsilon}) it holds that $\epsilon \Lambda = \frac{1}{4\sigma}$,
since after the $k$'th restart we have $\epsilon = \frac{1}{2^k 4 \sigma}$ and $\Lambda = 2^k$. An immediate consequence is that in\cref{line:price_update_scaling} it always holds for all $r \in \mathcal{R}$:
\begin{equation}
\epsilon \xi b_r \leq \frac{1}{4\sigma}.
\label{eq:bound_partial_usage}\end{equation}
Let us first fix some notation for the remainder of this section:
\begin{itemize}
    \item[(i)] We denote $K^* := \lceil \log \lambda^* \rceil$.
    \item[(ii)] We denote by $K$ the number of restarts of the algorithm (i.e. times the if-statement in \cref{line:check_price_bound} is satisfied). Further, we denote by $t_1\leq ... \leq t_K$ the indices of the phases in which the restarts occur. Note that these are not necessarily distinct. 
    \item[(iii)] For a phase $t$ we write $\epsilon^{(t)}$ for the $\epsilon$-value with which it was completed successfully. As usual, we write $x^{(t)} := \frac{1}{t} \sum_{p=1}^t s^{(p)}$ for the current solution after phase $t$. 
\end{itemize}

By construction of the algorithm, we have that after phase $t$ was completed
\begin{equation} y_r^{(t)} = \exp \left( \sum_{p=1}^t \epsilon^{(p)} s^{(p)}_r \right) \label{eq:price_with_epsilon_p}.\end{equation}
This allows us to formulate the following bound (analogous to the previous section) on the maximum entry of the current solution in terms of the price vector $y^{(t)}$. 
\begin{lemma}
The maximum entry of the current solution at the end of phase $t$ can be bounded in the following way:
\begin{equation}
\norm{x^{(t)}}_{\infty} \leq \frac{1}{\epsilon^{(t)} t} \log \norm{y^{(t)}}_1.
\label{eq:usage_bound}\end{equation}
\end{lemma}
\begin{proof}
Note that $\epsilon^{(p)}$ is decreasing for $p=1,...,t$. Utilizing \cref{eq:price_with_epsilon_p} we get for all $r \in \mathcal{R}$:
\[
    x^{(t)}_r  = \frac{1}{t} \sum_{p=1}^t s^{(p)}_r  
     = \frac{1}{t \epsilon^{(t)}} \log \exp\left( \epsilon^{(t)} \sum_{p=1}^t s^{(p)}_r\right) 
     \leq \frac{1}{t \epsilon^{(t)}} \log \exp\left( \sum_{p=1}^t \epsilon^{(p)} s^{(p)}_r\right) 
     = \frac{1}{t \epsilon^{(t)}} \log y_r^{(t)} 
     \leq \frac{1}{t \epsilon^{(t)} } \log \norm{y^{(t)}}_1.
\]
Since this bound is uniform for all $r \in \mathcal{R}$, it proves the desired result.
\end{proof}

At first glance, it is not clear that \cref{alg:fast_constant_factor_approximation} terminates. A phase might be restarted over and over again. In the following, we give a precise upper bound on the number of total restarts.
\begin{lemma}
\label{lem:bound_restarts}
It holds that $K \leq K^*$, i.e. the total number of restarts $K$ is bounded by $K^*$.
\end{lemma}
\begin{proof}
The proof is conducted along the lines of the proof of \cref{lem:rs_bound_price_increase}.
If $K < K^*$, then there is nothing to prove. Otherwise, let $t$ be the index of the phase where the $K^*$-th restart occurs. After the restart we have $\epsilon = \frac{1}{2^{K^*} 4 \sigma} \leq \frac{1}{4\sigma \lambda^*}$. Further, by our price bound in \cref{line:check_price_bound}, we know that for the prices in the previous phase $\norm{y^{(t-1)}}_1 \leq m \exp \left( t-1\right)$ holds (note that this holds also in the case $t=1$). Now we bound the increase of the sum of the prices in phase $t$.  In the following, we pretend that the if-statement in \cref{line:check_price_bound} does not exist and that phase $t$ is completed. We estimate $\norm{y^{(t)}}_1$ at the end. If it is below $m \exp (t)$, then we know that no restart occurred in phase $t$. Again, we fix the notation that  $y^{(t,j)}$ denotes the price vector $y$ after the $j$-th iteration of phase $t$, $\xi^{(t,j)}, b^{(t,j)}$ the $\xi$- and $b$-value in that iteration and $C(t,j)$ the currently considered customer in that iteration. We get the following estimate:  
\begin{flalign*}
    \sum_{r \in \mathcal{R}} y_r^{(t,j)} &  =\sum_{r \in \mathcal{R}} y_r^{(t,j-1)}\exp (\epsilon \xi^{(t,j)}b_r^{(t,j)}) \\
    & \leq \sum_{r \in \mathcal{R}} y_i^{(t,j-1)}\left( 1+ \frac{e^{1/(4\sigma)}-1}{1/(4\sigma)}\epsilon \xi^{(t,j)}  b_r^{(t,j)}\right)  & \text{ convexity of $\exp$, using \eqref{eq:bound_partial_usage}} \\
    & \leq \sum_{r \in \mathcal{R}} y_r^{(t,j-1)}  + \frac{e^{1/(4\sigma)}-1}{\lambda^*} \xi^{(t,j)} \inp{y^{(t,j-1)}}{b^{(t,j)}} & \epsilon \leq \frac{1}{4\sigma \lambda^*} \\
    & \leq \sum_{r \in \mathcal{R}} y_r^{(t,j-1)}  + \frac{\sigma (e^{(1/4\sigma)} - 1)}{\lambda^*} \xi^{(t,j)}  opt_{C(t,j)}(y^{(t,j-1)}) & \text{ approximation of the block solver} \\
    & \leq \sum_{r \in \mathcal{R}} y_r^{(t,j-1)}  + \frac{\sigma (e^{(1/4\sigma)} - 1)}{\lambda^*} \xi^{(t,j)} opt_{C(t,j)}(y^{(t)}). & \text{ prices increase}
\end{flalign*}
Applying the inequality recursively and using weak duality: $\sum_{C \in \mathcal{C}} opt_C(y^{(t)}) \leq \lambda^* \norm{y^{(t)}}_1$, as well as $\sum_{j: C(t,j) = C}\xi^{(t,j)} = 1$ for all $C$, one obtains (with $\kappa_t$ denoting the total number of oracle calls in phase $t$):
\begin{align*}
    \norm{y^{(t)}}_1 &\leq \norm{y^{(t-1)}}_1 + \frac{\sigma (e^{(1/4\sigma)} - 1)}{\lambda^*} \sum_{j=1}^{\kappa_t} \xi^{(t,j)} opt_{C^{(t,j)}}(y^{(t)}) \\
    & = \norm{y^{(t-1)}}_1 + \frac{\sigma (e^{(1/4\sigma)} - 1)}{\lambda^*} \sum_{C \in \mathcal{C}} opt_{C}(y^{(t)}) \\
    & \leq \norm{y^{(t-1)}}_1 + \frac{\sigma (e^{(1/4\sigma)} - 1)}{\lambda^*} \lambda^* \norm{y^{(t)}}_1 \\
    & = \norm{y^{(t-1)}}_1 + \sigma (e^{(1/4\sigma)} - 1) \norm{y^{(t)}}_1.
\end{align*}
Applying the bound $e^{(1/4\sigma)} \leq 1+\frac{1}{2\sigma}$ (as $\sigma \geq 1$), we get 
$\norm{y^{(t)}}_1 \leq \norm{y^{(t-1)}}_1 + \frac{1}{2}\norm{y^{(t)}}_1$ and thus 
\[ ||y^{(t)}||_1 \leq 2||y^{(t-1)}||_1 \leq e m \exp(t-1) = m\exp(t),\]
where we applied the bound $\norm{y^{(t-1)}}_1 \leq m \exp(t-1) $ (and $2 \leq e$).
This shows that phase $t$ will not be restarted after the $K^*$-th restart. Inductively, the same argument can be used to conclude that after the $K^*$-th restart no restart will occur anymore at all, which proves the claim.
\end{proof}
\begin{corollary}
\label{cor:bound_restarts}
Since $\lambda^* \leq \sigma m$, the number of restarts is in $\mathcal{O}(\log m)$.
\end{corollary}

With the previous statements, the approximation guarantee follows easily.
\begin{theorem}[Approximation guarantee]
\label{theo:approximation_guarantee}
The final solution $x^{(T)}$ satisfies 
\[ \norm{x^{(T)}}_{\infty} \leq 16 \sigma \lambda^*. \]
\end{theorem}
\begin{proof}
We know by \cref{lem:bound_restarts} that $\epsilon$ is halved at most $K^*$ many times and thus $\epsilon^{(T)} \geq \frac{1}{2^{K^*} 4 \sigma} \geq \frac{1}{\lambda^* 8 \sigma}$. Since it is guaranteed by \cref{line:check_price_bound} in the algorithm that $\norm{y^{(T)}}_1 \leq m \exp(T)$, we get with \cref{eq:usage_bound} that 
\[ \norm{x^{(T)}}_{\infty} \leq \frac{1}{\epsilon^{(T)} T}\log \norm{y^{(T)}}_1 \leq  \frac{ \log m}{\epsilon^{(T)}T} + \frac{1}{\epsilon^{(T)}} \leq 8 \sigma \lambda^* + 8 \sigma \lambda^* \leq 16 \sigma \lambda^*, \]
where we used $T = \lceil \log m \rceil$.
\end{proof}
This shows that \cref{alg:fast_constant_factor_approximation} is a $16\sigma$-approximation to the resource sharing problem. It remains to bound the running time. To this end, we distinguish between \emph{standard} and \emph{restricted} oracle calls. The first term refers to those for which the minimum in \cref{line:decide_xi_value} of \cref{alg:fast_constant_factor_approximation} is attained by $1-\alpha$. The second term refers to those in which the minimum in \cref{line:decide_xi_value} of \cref{alg:fast_constant_factor_approximation} is attained by $\frac{\Lambda}{\norm{b}_{\infty}}$.
It is easy to bound the number of standard oracle calls.
\begin{lemma}
\label{lem:standard_oracle_calls_bound}
The number of standard oracle calls in \cref{alg:fast_constant_factor_approximation} is in $\mathcal{O}(n \log m)$.
\end{lemma}
\begin{proof}
Clearly, in each phase, there are at most $n$ standard oracle calls. Since the total number of phases is  $T + K$, we have at most $n(T+K) = \mathcal{O}(n \log m)$ (by definition of $T$ and \cref{cor:bound_restarts}) many standard oracle calls.
\end{proof}

It remains to bound the number of restricted oracle calls. We prove a bound that depends on the indices $t_1 \leq ... \leq t_K$. The proof is analogous to the proof of \cref{lem:bound_number_of_oracle_calls_rs}.
\begin{lemma}
\label{lem:restricted_oracle_calls_bound}
The number of restricted oracle calls in \cref{alg:fast_constant_factor_approximation} is bounded by 
\[ 4 \sigma m\left(T + \sum_{i=1}^K t_i \right) + K.\]
\end{lemma}
\begin{proof}
The central observation is that in every restricted oracle call there is an $r \in \mathcal{R}$ such that $\xi = \Lambda/b_r$. Thus we have $\epsilon \xi b_r = \epsilon \Lambda = \frac{1}{4\sigma}$. This means that for every restricted oracle call there is a resource $r \in \mathcal{R}$ for which the resource price $y_r$ increases by $\exp(1/(4\sigma))$. Now we will bound the number of restricted oracle calls between the $(i-1)$ and $i$-th restart by $1+4m \sigma t_i$. Let us consider $i \in \{1,....,K\}$. For $r \in \mathcal{R}$ let $\kappa_r$ be the number of restricted oracle calls between the $(i-1)$-th and just before the $i$-th restart for which $y_r$ is increased by $\exp(1/(4\sigma))$. We can relate this quantity to $y_r$ in the following way: $y_r \geq \exp(\kappa_r/(4\sigma))$. Thus $\kappa_r \leq 4 \sigma \log y_r$.\\
Let $y^{(t_i)}$ denote the price vector in phase $t_i$, one oracle call before the $i$-th restart occurs (meaning that $\norm{y^{(t_i)}}_1 \leq m \exp (t_i)$ holds and is violated only in the next iteration).  We can bound the total number of restricted oracle calls between the $(i-1)$-th and $i$-th restart by:
\begin{flalign*}
1 + \sum_{r \in \mathcal{R}} \kappa_r  & \leq 1 + 4 \sigma \sum_{r \in \mathcal{R}} \log y^{(t_i)}_r & \text{ One oracle call that violates the bound } \\
& = 1+ 4 \sigma \log \prod_{r \in \mathcal{R}} y_r^{(t_i)} \\
& = 1+ 4 \sigma m \log \left( \prod_{r \in \mathcal{R}} y_r^{(t_i)}\right)^{\frac{1}{m}} \\
& \leq 1+ 4 \sigma m \log \frac{\norm{y^{(t_i)}}_1}{m} & \text{ Inequality between geometric and arithmetic mean} \\
& \leq 1+ 4 \sigma m \log \frac{m \exp(t_i)}{m} & \text{ bound on  $\norm{y^{(t_i)}}_1$} \\
& = 1+ 4 m \sigma t_i.
\end{flalign*}
This holds for every $i$-th restart for $i =1,...,K$. The very same argument can be applied to bound the number of restricted oracle calls after the $K$-th restart up to phase $T$. In this case, one does not need to add 1, because the bound $y^{(T)} \leq m \exp(T)$ is not violated. This proves the claim.
\end{proof}
At first, \cref{lem:restricted_oracle_calls_bound} does not seem to be very useful, as we only know $t_i \leq T$ and $K \in \mathcal{O}(\log m)$. A naive estimate would only give a bound of $\mathcal{O}(m \log^2 m)$ many restricted oracle calls, which would not yield any improvement.  However, intuitively, one would expect that if $\lambda^*\gg 1$, then the prices should increase fast, and the bound in \cref{line:check_price_bound} should be violated already after few phases. This is exactly the reason why this approach works. Indeed, we can provide upper bounds on the $t_i$, which allow estimating $\sum_{i=1}^K t_i$ by a geometric series resulting in $\sum_{i=1}^K t_i \in \mathcal{O}(\log m)$. This is the statement of the following lemma.

\begin{lemma}
\label{lem:ti_bound}
The following bound on $t_i$ is valid for all $i \leq K^*-3-\lceil \log \sigma\rceil$:
\[ t_i \leq  1 + 2^{i-K^* + 3} \sigma \log m.\]
Therefore, 
\[ \sum_{i=1}^K t_i \leq (5 + \lceil \log \sigma \rceil) \lceil \log m \rceil + \lceil \log \sigma m \rceil . \]
\end{lemma}
\begin{proof}
If $t_i = 1$ then there is nothing to prove. Otherwise, when the $i$-th restart occurs in phase $t_i$, the value of $\epsilon$ is reduced to $\epsilon = \frac{1}{2^i 4 \sigma}$. Therefore, we know that phase $t_i-1$ ($\geq 1$)  was completed successfully with an $\epsilon^{(t_i - 1)} \geq \frac{1}{2^{i-1} 4 \sigma}$. Since it was completed successfully, we know that $\norm{y^{(t_i-1)}}_1 \leq m \exp(t_i -1)$ held. Now we consider the  current solution $x^{(t_i-1)}$ after phase $t_i-1$ was completed. Of course $\lambda^* \leq \norm{x^{(t_i-1)}}_{\infty}$ must hold. The bound \eqref{eq:usage_bound} states for the solution $x^{(t_i -1)}$ that 
\[\norm{x^{(t_i -1)}}_{\infty} \leq \frac{1}{(t_i -1) \epsilon^{(t_i -1)}} \log \norm{y^{(t_i-1)}}_1 \leq 2^{i-1} 4 \sigma \frac{\log \norm{y^{(t_i-1)}}_1}{t_i - 1}.\]
Combining the lower bound on the left-hand side and the upper bound on the right-hand side we can deduce 
\[ \lambda^* \leq 2^{i-1} 4 \sigma \frac{\log m + (t_i -1)}{t_i - 1},\]
or equivalently 
\[ \frac{\lambda^*}{2^{i-1} 4 \sigma} - 1 \leq \frac{\log m }{t_i -1}.\]
Recall that by definition $K^* = \lceil \log_2 \lambda^* \rceil$ and therefore we get the estimate
\[ \frac{2^{K^* -1}}{ 2^{i-1}4 \sigma} - 1 \leq \frac{\log m}{t_i -1}. \]
We can simplify it slightly to 
\[  \frac{2^{K^* - i - 2}}{\sigma} - 1 \leq \frac{\log m }{t_i -1}. \]
If $i \leq K^* -3 - \lceil \log \sigma \rceil$ we have that $\frac{2^{K^* - i - 2}}{\sigma} \geq 2$ and thus $\frac{2^{K^* - i - 3}}{\sigma} \leq \frac{2^{K^* - i - 2}}{\sigma} - 1$. This leads to
\[ \frac{2^{K^* - i - 3}}{\sigma} \leq \frac{\log m }{t_i -1}, \]
which finally allows us to conclude the desired bound 
\[ t_i \leq 1+ 2^{i- K^*+3} \sigma \log m . \]
Since we know that $K \leq K^* \leq \lceil \log \sigma m \rceil$, and $t_i \leq T$ for all $i$, we can estimate the sum as follows
\begin{flalign*}
    \sum_{i=1}^K t_i & \leq (3+\lceil \log \sigma \rceil) T + \sum_{i=1}^{K^*- 3 - \lceil \log \sigma \rceil} t_i & \text{ Estimate last summands by $T$} \\
    & \leq (3+\lceil \log \sigma \rceil) T +  \sum_{i=1}^{K^*- 3 - \lceil \log \sigma \rceil} (1+ 2^{i- K^*+3}\sigma \log m ) & \text{ Bound on $t_i$ above} \\
    & \leq (3+\lceil \log \sigma \rceil) T + K^* + \sigma \log m \sum_{i=1}^{K^*- 3 - \lceil \log \sigma \rceil} 2^{i- K^*+3} & \\
    & \leq  (3+\lceil \log \sigma \rceil) T + K^* + \sigma \log m \sum_{j=\lceil \log \sigma \rceil}^{\infty} 2^{-j} & \\
    & \leq (3+\lceil \log \sigma \rceil) T + K^* + 2 \log m & \\
    & \leq (5 + \lceil \log \sigma \rceil) \lceil \log m \rceil + \lceil \log \sigma m \rceil. & T = \lceil \log m\rceil , K^* \leq \lceil \log \sigma m \rceil .
\end{flalign*}
\end{proof}
 Combining this with \cref{lem:restricted_oracle_calls_bound}, we can get a good bound on the number of restricted oracle calls.
\begin{corollary}
The number of restricted oracle calls in \cref{alg:fast_constant_factor_approximation} is in $\mathcal{O}(m \log m)$.
\end{corollary}
\begin{proof}
 \cref{lem:ti_bound} shows $T + \sum_{i=1}^K t_i \in \mathcal{O}(\log m)$. The statement follows then immediately with \cref{lem:restricted_oracle_calls_bound}.
\end{proof}

We are now able to summarize the analysis of the algorithm in the following theorem.
\begin{theorem}
\cref{alg:fast_constant_factor_approximation} computes a solution $x^{(T)} \in \mathcal{X}$ to the resource sharing problem that satisfies 
\[ \norm{x^{(T)}}_{\infty} \leq  16 \sigma \lambda^* \]
and uses $\mathcal{O}((n+m) \log m)$ oracle calls.
\label{theo:fast_constant_factor_approximation}
\end{theorem}
\begin{remark}
\cref{alg:fast_constant_factor_approximation} works in the same asymptotic running time if it is restarted completely whenever the bound in \cref{line:check_price_bound} is violated. However, we described it in this way to show that this does not need to be done.
\end{remark}
To conclude the proof of the main theorem, \cref{theo:rs_main_theorem}, we run \cref{alg:fast_constant_factor_approximation} to obtain a primal solution $x^{(T)} \in \mathcal{X}$ that satisfies $\norm{x^{(T)}}_{\infty} \leq 16 \sigma \lambda^*$. Then we normalize the instance to obtain a new one $\tilde{\mathcal{X}} := \frac{1}{\norm{x^{(T)}}_{\infty}}$. We know for the optimum objective value $\tilde{\lambda}^*$ on this new instance that $\tilde{\lambda}^* \in [\frac{1}{16 \sigma}, 1]$. Thus we can run the core algorithm on this new instance (with parameters specified according to the proof of \cref{theo:rs_normalized_main_theorem} and rescaling the obtained primal solution) to prove the main theorem.

\section{Decreasing minimality on the two largest entries}
\label{sec:dec_min_two}
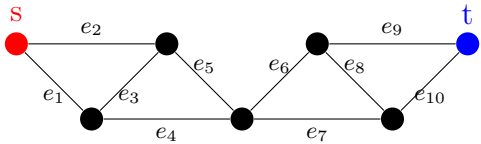
\begin{figure}[t]
    \centering
    \begin{tikzpicture}
    \draw node[circle, fill = black] (A) at (0,0) {};
    \draw node[circle, fill = red] (B) at (-1,1) {};
    \draw node[circle, fill = black] (C) at (1,1) {};
    \draw node[circle, fill = black] (D) at (2,0) {};
    \draw node[circle, fill = black] (E) at (3,1) {};
    \draw node[circle, fill = black] (F) at (4,0) {};
    \draw node[circle, fill = blue] (G) at (5,1) {};
    \node at (-1,1.4) {\large \textcolor{red}{s}};
    \node at (5,1.4) {\large \textcolor{blue}{t}};
    \draw (A) --  (B) node[midway,below] {$e_1$};
    \draw (B) -- (C) node[midway, above] {$e_2$};
    \draw (A) -- (C) node[midway, below] {$e_3$};
    \draw (A) -- (D) node[midway, below] {$e_4$};
    \draw (C) -- (D) node[midway, above] {$e_5$};
    \draw (D) --  (E) node[midway, above] {$e_6$};
    \draw (D) -- (F) node[midway, below] {$e_7$};
    \draw (E) -- (F) node[midway, above] {$e_8$};
    \draw (E) -- (G) node[midway, above] {$e_9$};
    \draw (F) -- (G) node[midway, below] {$e_{10}$};
    \end{tikzpicture}
    \caption{A network with a cut vertex. One could consider the problem of computing a maximum $s$-$t$ flow that minimizes the maximum flow that is sent along any edge. Then this maximum over $\mathcal{R}_1 = \{e_1,...,e_5\}$ and $\mathcal{R}_2 = \{e_6,...,e_{10} \}$ is optimized independently with our algorithm.}
    \label{fig:network_with_cut_vertex}
\end{figure}

In this section, we discuss the implications of local weak duality. In particular, we prove that our algorithm computes solutions that are close to decreasingly minimal on the two largest entries. Before that, let us consider a simple application, which is the \emph{product case}.
Assume that the instance of the resource sharing problem consists of multiple independent parts, that is, there exists a (perhaps unknown) partition of the resources $\mathcal{R} = \mathcal{R}_1 \cup ... \cup \mathcal{R}_k$ on which the customers act "independently". More precisely, we assume that each of the convex sets $X_C$ can be decomposed into a product $X_C = X_1^{(C)} \times ... \times X_K^{(C)}$ where $X_i^{(C)} \subseteq \mathbb{R}^{|\mathcal{R}_i|}_{\geq 0}$ for $i = 1,...,k$. Let us write $\mathcal{X}_i := \sum_{C \in \mathcal{C}} X_i^{(C)}$ for $i = 1,...,k$. Note that we have $\mathcal{X} = \mathcal{X}_1 \times ... \times \mathcal{X}_k$. Let $C \in \mathcal{C}$ and $y^{(C)} \in \mathbb{R}^m_{\geq 0}$. It is clear that the linear minimization problem decomposes into $k$ separate optimization problems in the following way:
\[opt_C(y^{(C)}) = \min_{x^{(C)} \in X_C} \inp{y^{(C)}}{x^{(C)}} = \sum_{i=1}^k \min_{x^{(C)}_i \in X^{(C)}_i} \inp{y^{(C)}|_{\mathcal{R}_i}}{x^{(C)}_i},  \]
where $y^{(C)}|_{\mathcal{R}_i}\in \mathbb{R}^{\mathcal{R}_i}_{\geq 0}$ denotes the restriction of $y^{(C)}$ to the entries in $\mathcal{R}_i$. If the block solver is exact, we get for every $j=1,...,k$ that the local objective value is bounded by
\begin{align*} 
\sum_{r \in \mathcal{R}_j} y^{(C)}_r f_C(y^{(C)})_r & = opt_C(y^{(C)}) - \sum_{r \in \mathcal{R}\setminus \mathcal{R}_j} y^{(C)}_r f_C(y^{(C)})_r \\
& \leq opt_C(y^{(C)}) - \sum_{i=1, i\neq j}^k \min_{x^{(C)}_i \in X^{(C)}_i} \inp{y^{(C)}|_{\mathcal{R}_i}}{x^{(C)}_i} \\
& = \min_{x^{(C)}_j \in X^{(C)}_j} \inp{y^{(C)}|_{\mathcal{R}_j}}{x^{(C)}_j}.
\end{align*}
Now choose a $y^{(C)} \in \mathbb{R}^m_{\geq 0}$ for every customer $C$ and let $\bar{y} := (\max_{C \in \mathcal{C}} y^{(C)}_r)_{r \in \mathcal{R}}$ denote their pointwise maximum. We want to prove that $\mathcal{X}$ satisfies local weak duality w.r.t. $\mathcal{R}_j$ and $\min_{x_j \in \mathcal{X}_j} \norm{x_j}_{\infty}$ for every $j=1,...,k$. This is easily verified as follows:
\begin{align*}
    \sum_{C \in \mathcal{C}} \sum_{r \in \mathcal{R}_j} y^{(C)}_r f_C(y^{(C)})_r & \leq \sum_{C \in \mathcal{C}}\min_{x^{(C)}_j \in X^{(C)}_j} \inp{y^{(C)}|_{\mathcal{R}_j}}{x^{(C)}_j} \\
    & \leq \sum_{C \in \mathcal{C}} \min_{x^{(C)}_j \in X^{(C)}_j} \inp{\bar{y}|_{\mathcal{R}_j}}{x^{(C)}_j} \\
    & = \min_{x_j \in \mathcal{X}_j} \inp{\bar{y}|_{\mathcal{R}_j}}{x_j} \\
    & \leq \min_{x_j \in \mathcal{X}_j} \norm{x_j}_{\infty} \sum_{r \in \mathcal{R}_j}\bar{y}_r.
\end{align*}
Therefore, in this case, $\mathcal{X}$ satisfies local weak duality w.r.t. $\mathcal{R}_j$ and $\min_{x_j \in \mathcal{X}_j} \norm{x_j}_{\infty}$ for every $j =1,...,k$. So, indeed, according to the main theorem, \cref{theo:rs_main_theorem}, every independent part of the instance is optimized separately in the sense of the following theorem (which is an immediate consequence of \cref{theo:rs_main_theorem}).
\begin{theorem}
\label{theo:rs_product_structure}
Let $\mathcal{X}$ be an instance of the resource sharing problem with an (unknown) product structure as described above. Assume that we have an exact block solver, i.e. $\sigma = 1$. Then for every $\delta \in  (0,1]$ our algorithm computes a solution $x \in \mathcal{X}$ satisfying
\[ \max_{r \in \mathcal{R}_i} x_r \leq  \min_{x_i \in \mathcal{X}_i} \norm{x_i}_{\infty} + \delta \lambda^*\]
for all $i =1,...,k$, using $\mathcal{O}\left( \frac{n+m}{\delta^2}\log m\right)$ oracle calls.
\end{theorem}
 A natural example of instances with a product structure can be found in network flow problems where the network has a cut vertex, e.g. \cref{fig:network_with_cut_vertex}.
\begin{remark}
In the product case, a similar result can be deduced for approximate block solvers, if we can assume that the approximation guarantee holds locally for every independent part.
\end{remark}

In the following, we want to prove that any instance $\mathcal{X}$ with an exact block solver satisfies local weak duality w.r.t. to all but one coordinate and the value of the second-highest entry in the decreasingly minimal solution. To do so, we need a helper lemma about the structure of decreasingly minimal solutions in convex sets.
It is possible to use the convexity of the set $\mathcal{X}$ to make a statement about the explicit coordinates in which the decreasingly minimal element is maximal and relate it to other elements in $\mathcal{X}$ as follows.
\begin{lemma}
Let $X \subseteq \mathbb{R}^m_{\geq 0}$ be a non-empty closed convex set. Let $\lambda \in X$ be the decreasingly minimal element and let $I := \{ i \in \{1,...,m\} : \lambda_i = \norm{\lambda}_{\infty} \}$ be the subset of all indices for which $\lambda$ is maximal. Then the following property is valid:
\[\forall x \in X \qquad \text{ either } \qquad  \forall i \in I: x_i = \lambda_i  \qquad \text{ or } \qquad \exists j \in I : x_j > \lambda_j .\]
\label{lem:lex_min_support}
\end{lemma}
\begin{proof}
Assume that this is not the case. Then there exists an $x \in X$ such that $x_i \leq \lambda_i$ for all $i \in I$ and there is a $j \in I$, such that $x_j < \lambda_j$. We want to construct an element
\[ z = tx+(1-t) \lambda \]
that satisfies $z_i \leq \norm{\lambda}_{\infty}$ for all $i \in \{1,...,m\}$ and $z_k < \norm{\lambda}_{\infty}$ for all $k \in (\{1,...,m\}\setminus I) \cup \{j\}$. Then $z$ has fewer maximal entries than $\lambda$ and $z <_{dec} \lambda$ follows, which is a contradiction to the decreasing  minimality of $\lambda$. Let $\varepsilon > 0$ be such that $\forall k \in \{1,...,m \} \setminus I : \norm{\lambda}_{\infty} - \lambda_k > \varepsilon$. We claim that $t := \frac{\varepsilon}{\norm{x}_{\infty}}$ satisfies the properties claimed above. Since $x_i \leq \norm{\lambda}_{\infty}$ for all $i \in I$ we have that $z_i \leq \norm{\lambda}_{\infty}$. Further, for the special index $j$ we get 
\[ z_j = tx_j + (1-t) \lambda_j < \norm{\lambda}_{\infty} \]
because $x_j < \norm{\lambda}_{\infty}$. Finally, for $k \notin I$, we have
\[ z_k = \frac{\varepsilon}{\norm{x}_{\infty}} x_k + \left(1-\frac{\varepsilon}{\norm{x}_{\infty}}\right) \lambda_k \leq \varepsilon + \lambda_k < \norm{\lambda}_{\infty} \]
by definition of $\varepsilon$. 
\end{proof}
\begin{remark}
\label{rem:useful_for_two_in_rs}
The simplest application of this Lemma is when $I$ consists of a single index, i.e. $I = \{i\}$. Then the Lemma states that $\forall x \in X: x_i \geq \lambda_i$.
\end{remark}

\begin{lemma}
\label{lem:monotone_weak_duality_second_entry}
Let $\mathcal{X}$ be an instance of the resource sharing problem with $\sigma =1$. Let $\lambda \in \mathcal{X}$ be the decreasingly minimal element. Assume that $(\lambda_{\downarrow})_1 > (\lambda_{\downarrow})_2$ holds. Let $r^* \in \mathcal{R}$ be the coordinate of the unique maximum entry, i.e. $\lambda_{r^*} = (\lambda_{\downarrow})_1$. Then $\mathcal{X}$ satisfies local weak duality w.r.t. $S:= \mathcal{R}\setminus \{r^*\}$ and $\mu := (\lambda_{\downarrow})_2$.
\end{lemma}
\begin{proof}
Since $r^*$ is the unique coordinate in which $\lambda$ is maximal, we know by \cref{lem:lex_min_support} that $\min_{x \in \mathcal{X}} x_{r^*} = \lambda_{r^*}$. Since $\mathcal{X} = \sum_{C \in \mathcal{C}} X_C$ it also follows that $\sum_{C \in \mathcal{C}} \min_{x^{(C)} \in X_C} x^{(C)}_{r^*} = \lambda_{r^*}$. Now let $(y^{(C)})_{C \in \mathcal{C}}\subseteq \mathbb{R}^m_{\geq 0}$ be an arbitrary collection of price vectors. Independently of the previous argument, one can exploit the definition of $S$, and that the block solver is exact, to get the estimate
\begin{align*} \sum_{r \in S} y^{(C)}_r f_C(y^{(C)})_r & =  opt_C(y^{(C)}) - y^{(C)}_{r^*} f_C(y^{(C)})_r \\ 
& \leq opt_C(y^{(C)}) - y^{(C)}_{r^*} \min_{x^{(C)} \in X_C} x^{(C)}_{r^*}.\end{align*}
We want to argue that the right-hand side is monotone increasing in the price vector. For that, take an upper bound $y\geq y^{(C)}$. Let $v^{(C)} \in X_C$ that attains $\inp{y}{v^{(C)}} = opt_C(y)$. We have
\begin{align*}
    opt_C(y) - y_{r^*} \min_{x^{(C)} \in X_C} x^{(C)}_{r^*} & = \inp{y}{v^{(C)}} - y_{r^*} \min_{x^{(C)} \in X_C} x^{(C)}_{r^*}\\
    & = \sum_{r \in S} y_r v^{(C)}_r + y_{r^*} \underbrace{(v^{(C)}_{r^*} - \min_{x^{(C)} \in X_C} x^{(C)}_{r^*})}_{\geq 0} \\
    & \geq \sum_{r \in S} y^{(C)}_r v^{(C)}_r + y^{(C)}_{r^*} (v^{(C)}_{r^*} - \min_{x^{(C)} \in X_C} x^{(C)}_{r^*}) \\
    & = \inp{y^{(C)}}{ v^{(C)}} - y^{(C)}_{r^*} \min_{x^{(C)} \in X_C} x^{(C)}_{r^*} \\
    & \geq opt_C(y^{(C)}) - y^{(C)}_{r^*}\min_{x^{(C)} \in X_C} x^{(C)}_{r^*}.
\end{align*}
Thus for a collection of price vectors $(y^{(C)})_{C \in \mathcal{C}} \subset \mathbb{R}^m_{\geq 0}$ and their pointwise maximum $\bar{y} = (\max_{C \in  \mathcal{C}}y^{(C)}_r)_{r \in \mathcal{R}}$, we get
\[ \sum_{C \in \mathcal{C}} \sum_{r \in S} y^{(C)}_r f_C(y^{(C)})_r  \leq \sum_{C \in \mathcal{C}} opt_C(\bar{y}) - \bar{y}_{r^*}\sum_{C \in \mathcal{C}} \min_{x^{(C)} \in X_C} x^{(C)}_{r^*} = \min_{x \in \mathcal{X}} \inp{\bar{y}}{x} - \bar{y}_{r^*} (\lambda_{\downarrow})_1.\]
Since $\lambda \in \mathcal{X}$ is a feasible solution we get the estimate
\[ \min_{x \in \mathcal{X}} \inp{\bar{y}}{x} \leq \inp{\bar{y}}{\lambda} \leq  \left( \sum_{r \in S} \bar{y}_r \right) (\lambda_{\downarrow})_2 + \bar{y}_{r^*} (\lambda_{\downarrow})_1. \]
This inequality can be inserted into the previous one to finally obtain (inserting the definition of $\mu$)
\[ \sum_{C \in \mathcal{C}}  \sum_{r \in S} y^{(C)}_r f_C(y^{(C)}) \leq \mu \sum_{r \in S} \bar{y}_r.\]
Therefore, $\mathcal{X}$ satisfies local weak duality w.r.t. $S$ and $\mu$.
\end{proof}

Note that if $(\lambda_{\downarrow})_1 = (\lambda_{\downarrow})_2$ \cref{cor:dec_min_two} is trivial due to primal convergence. In the case $(\lambda_{\downarrow})_1 > (\lambda_{\downarrow})_2$, \cref{cor:dec_min_two} follows due to local weak duality with an application of the main theorem. \\\\

One might conjecture that an analogous version to \cref{cor:dec_min_two} is valid also for the third-highest entry and so on (meaning that the computed solution $x$ satisfies $(x_{\downarrow})_3 \leq (\lambda_{\downarrow})_3 + \delta \lambda^*$ etc.). We prove in the following that this is not the case in general. To do so, we need a helper lemma, which states that also dual convergence of the non-average price vectors $y^{(t)}$ is achieved in the core algorithm (although at a slower rate than the average price). We state it only in the case $\lambda^* = 1$, as we only apply this special case.
\begin{lemma}
Assume that $\lambda^* = 1$ and $\sigma = 1$. For every  $\delta \in (0,1]$ and every $\epsilon, T$ that satisfy $0 < \epsilon < \frac{1}{4} \delta^2$ and $T \geq \log m / \epsilon^2$  there is a $T' \geq (1-\delta) T$, such that the core algorithm run with parameters $\epsilon, T$ computes a price vector $y^{(T')}$ that satisfies 
\[ \Theta_{T'} \geq (1- \delta). \]
\label{coro:exist_high_cost_sol_late_it}
\end{lemma}
\begin{proof}
By assumption $\lambda^* =1$ and thus $\Theta_t \leq 1$ for all phases $t=1,...,T$. Now assume that the statement does not hold, i.e. $\Theta_t \leq (1- \delta)$ for all $t \geq (1-\delta)T$. Then we have the following bound:
\begin{equation} \sum_{t=1}^T \Theta_t \leq (1-\delta) T + \delta T (1-\delta) = T - \delta^2 T,  \label{eq:late_phase_ubound}\end{equation}
as the first $(1-\delta)T$ summands can be bounded by $1$ and the remaining $\delta T$ summands by $(1-\delta)$. Using the bound on the average dual value ( \cref{theo:primal_dual_error}) we also have:
\begin{equation} \frac{1}{T}\sum_{t=1}^T \Theta_t \geq \frac{1-\eta \lambda^*}{1+\epsilon}\left( 1 - \epsilon \right) \geq \frac{1-2\epsilon}{1+\epsilon} (1-\epsilon) \geq (1-3\epsilon) (1-\epsilon) \geq 1-4\epsilon. \label{eq:late_phase_lbound} \end{equation}
Dividing \eqref{eq:late_phase_ubound} by $T$ and combining it with \eqref{eq:late_phase_lbound} one gets
\[ 1-\delta^2 \geq 1-4\epsilon.\]
This is a contradiction to the assumption that $\epsilon < \frac{1}{4} \delta^2$.
\end{proof}
In the following, we construct an example that is "as bad as possible". We will, for given parameters $\epsilon$ and $T$ construct an adversarial instance on which the core algorithm terminates with a solution where every entry is "close" to $\lambda^*$, even though the decreasingly minimal solution $\lambda$ satisfies $(\lambda_{\downarrow})_3 = 0$. This is done for arbitrary $m \geq 3$. We show it in the following theorem.
\begin{theorem}
\label{theo:rs_all_entries_large}
Let $m \geq 3, \delta \in (0,\frac{1}{8}]$ and $K \in \mathbb{N}$. Then for any parameter $\epsilon$ that satisfies $0 < \epsilon < \frac{1}{4K}\delta^2$ there is an instance of the resource sharing problem which has the following properties:
\begin{itemize}
    \item The decreasingly minimal solution $\lambda$ is given by $\lambda = (1,1,0,...,0)$.
    \item The core algorithm (\cref{alg:mrv_resource_sharing_core}) run with parameters $\epsilon$,  $T \in \left\{ \left\lceil \frac{\log m }{\epsilon^2} \right\rceil,..., \left\lceil \frac{K \log m}{\epsilon^2}\right\rceil\right\}$ and an exact block solver returns a solution $x^{(T)}$ that satisfies:
    \[  \min_{r \in \mathcal{R}} x^{(T)}_r \geq 1 - 2 \delta.\]
\end{itemize}
\end{theorem}
Thus even if all apart from two entries vanish in the decreasingly minimal solution, the algorithm may compute a solution where every entry is close to $\lambda^*$. One may interpret this as saying that the algorithm computes a "worst possible" solution under the restriction that the maximum usage is close to $\lambda^*$.
\begin{proof}
We consider an instance that consists of a single customer (and depends on $\epsilon$). The set of feasible solutions is given as a convex hull of two  vertices $ \mathcal{X} = conv(\{\lambda, v\})$ where 
\[\lambda := (1,1,0,...,0), \qquad v := \left(\frac{5}{4},0,m^{\epsilon^{-1/2}},m^{\epsilon^{-1/2}},...,m^{\epsilon^{-1/2}}\right).\]
It is easy to see that $\lambda$ is the decreasingly minimal element in $\mathcal{X}$. Instead of analyzing the behavior of the core algorithm explicitly on a per-iteration basis we make a more abstract argument that is based on dual convergence of the algorithm. The general idea is that we want to use \cref{coro:exist_high_cost_sol_late_it} to argue that in a late iteration $T'\in \{\lceil(1-\delta)T\rceil, ...,T\}$ of the algorithm the price $y_i^{(T')}$ for $i\in \{3,...,m\}$ must be "high", otherwise, it would be possible to construct a cheap solution. This can then be used to argue that $x^{(T)}_i$ must also be high (i.e. close to 1). \\
Let us formally state the argument. The returned solution $x^{(T)}$ can be written as a convex combination $x^{(T)} = \xi^{(T)} v + (1-\xi^{(T)}) \lambda$.
We know by primal convergence of the core algorithm, \cref{theo:primal_dual_error}, that \[(x^{(T)}_{\downarrow})_1 = \norm{x^{(T)}}_{\infty} \leq\frac{\log m}{\epsilon T}+ \frac{1+\epsilon}{1-\eta \lambda^*} \lambda^* \leq \epsilon + \frac{1+\epsilon}{1-2\epsilon}= 1+\epsilon+\frac{3\epsilon}{1-2\epsilon} \leq 1+5\epsilon < 2.\] Therefore, we can bound the fraction with which $v$ is taken into account by $ \xi^{(T)} \leq 2m^{-\epsilon^{-1/2}}$. Let us denote $\alpha := 2m^{-\epsilon^{-1/2}}$.
Now the first step is to prove that for all phases $t$, the price of the second resource is at most by a factor of 2 smaller than that of the first, i.e.
\begin{equation}\forall t = 1,...,T : 2 y_2^{(t)} \geq y_1^{(t)} \label{eq:price_at_most_twice} \end{equation}
holds. This can be seen in the following way: At the end of each phase $t$, the current element can be written as a convex combination $x^{(t)} = \xi^{(t)} v + (1-\xi^{(t)}) \lambda$. Observe that the price of resources with index $i=3,...,m$ at the end of phase $t$ is given as $y^{(t)}_i = \exp(\epsilon t \xi^{(t)} m^{\epsilon^{-1/2}})$. By the previous observation, we know $y_i^{(T)} \leq \exp(\epsilon T \alpha m^{\epsilon^{-1/2}})$. This can be used to bound $t \xi^{(t)} \leq T \alpha$. Implications are the following inequalities: 
\[ y_1^{(t)} = \exp\left( \epsilon t \left(\frac{5}{4} \xi^{(t)} + (1-\xi^{(t)}) \right)\right) = \exp\left( \epsilon t + \frac{1}{4}\epsilon t \xi^{(t)}\right)\leq \exp(\epsilon t) \exp\left(\frac{1}{4}\epsilon T \alpha\right) \]
and 
\begin{equation} y_2^{(t)} = \exp\left(\epsilon t \left(1-\xi^{(t)}\right)\right)\geq \exp(\epsilon t) \exp(-\epsilon T \alpha). \end{equation}
Thus $y_2^{(T)} \exp(\frac{5}{4} \epsilon T \alpha )\geq y_1^{(T)}$, i.e. the claim is satisfied if $\frac{5}{4} \epsilon T \alpha \leq \log 2$. By assumption, $\epsilon$ is small enough such that $\frac{1}{\sqrt{\epsilon}} \geq 1+\log(5/2) + \log(K/\epsilon)$ holds. This allows deducing, using the upper bound on $T$, that
\[ \frac{5}{4} \epsilon \alpha T  \leq \frac{5}{4}\epsilon \alpha \left( \frac{K\log m}{\epsilon^2} + 1\right) = \frac{5}{2m^{1/\sqrt{\epsilon}}} \left(\frac{K\log m}{\epsilon} + \epsilon \right) \leq \frac{\log m}{m} + \frac{\epsilon^2}{Km} \leq  \log 2,\]
which holds for $m \geq 3$.\\
The second step is to use \cref{coro:exist_high_cost_sol_late_it}. It states that there is a phase $T' \in \{(1-\delta)T,...,T\}$ such that $\Theta_{T'} \geq (1-\delta)$. This means especially that the (normalized) objective value of choosing solution $v$ in phase  $T'$ is at least $1-\delta$. So
\[(1-\delta) \sum_{j=1}^m y_j^{(T')} \leq \inp{y^{(T')}}{v} =  \frac{5}{4} y_1^{(T')} + m^{\epsilon^{-1/2}}\sum_{j=3}^m y_j^{(T')}. \]
By \eqref{eq:price_at_most_twice} we know that $\sum_{j=1}^m y_j^{(t)}\geq y_1^{(t)}+ y_2^{(t)} \geq  \frac{3}{2}y_1^{(t)}$ for all phases $t=1,...,T$. Further, by construction of $\mathcal{X}$, it is easy to see that $y_1^{(t)} \geq \exp(\epsilon t)$ must hold after each phase $t$. This can be used to see that (by inserting the inequality above):
\begin{align*}
    m^{\epsilon^{-1/2}}\sum_{j=3}^my_j^{(T')} & \geq \left(\frac{3}{2} - \frac{5}{4}-\frac{3}{2}\delta\right)y_1^{(T')} \\ & \geq \left(\frac{1}{4}-\frac{3}{2}\delta\right) \exp(\epsilon T') \\ & \geq \left(\frac{1}{4}-\frac{3}{2}\delta\right) \exp(\epsilon (1-\delta) T).
\end{align*}  
Using that by construction of $\mathcal{X}$, always $y_3^{(t)} = ... = y_m^{(t)}$ holds, we get that for all $j=3,...,m$ it holds $y_j^{(T')} \geq m^{-\epsilon^{-1/2} -1}\left(\frac{1}{4}-\frac{3}{2}\delta\right) \exp(\epsilon (1-\delta) T)$.
This shows that at termination we have for all $j=3,...,m$
\begin{align*}  x^{(T)}_j & = \frac{1}{\epsilon T} \log y_j^{(T)} \\
& \geq \frac{1}{\epsilon T} \log y_j^{(T')} \\
& \geq (1-\delta) +\frac{1}{\epsilon T} \left(\log\left(\frac{1}{4}-\frac{3}{2}\delta\right)- (1+\epsilon^{-1/2}) \log m \right) \\
& \geq (1-\delta) + \frac{\epsilon}{\log m } \log\left(\frac{1}{4}-\frac{3}{2}\delta\right)- (\epsilon + \epsilon^{1/2}) \\
& \geq 1-\delta + \frac{\delta^2}{4 \log m} \log \left( \frac{1}{16}\right) -  \frac{\delta^2}{4} - \frac{1}{2}\delta \\
& \geq 1- \delta - \delta^2  - \frac{\delta^2}{4} -\frac{1}{2} \delta \\
& \geq 1-2\delta,
\end{align*}
where we used $\delta \leq \frac{1}{8}$, $\epsilon \leq \delta^2/4$, $T \geq \lceil \log m /\epsilon^2 \rceil$ and the previous inequality to bound $y_i^{(T')}$ from below. By construction of the instance, we have $x_1^{(T)} \geq 1$, and by \eqref{eq:price_at_most_twice}: $x_2^{(T)} \geq x_1^{(T)} - \frac{\log 2}{\epsilon T} \geq 1-\epsilon$. This proves $\min_{r \in \mathcal{R}} x_r^{(T)} \geq 1-2\delta$.
\end{proof}

\section{A tight lower bound on the number of required oracle calls}
\label{sec:lower_bound_iterations}
In this section, we prove \cref{theo:lower_bound_iterations_rs}. Klein and Young \cite{klein_young_lower_bound_iterations}  provide the following lower bound on the number of oracle calls for any Dantzig-Wolfe-type algorithm in the context of fractional packing.
\begin{theorem}[Lower bound on the number of oracle calls, \cite{klein_young_lower_bound_iterations} Corollary 2.1]
\label{theo:klein_young_bound}
For every $\gamma \in (0,1/2)$, there exist positive constants $k_{\gamma}, c_{\gamma} > 0$, such that  $\forall m,k \geq k_{\gamma}$ and $\rho \geq 2$ there exists a packing input $(A, b, X_P)$, with $A \in \mathbb{R}^{m\times k}$ and width in $\mathcal{O}(\rho)$ such that:\\
For every $\delta \in (0,1/10)$ every Dantzig-Wolfe-type algorithm requires at least 
\[ c_{\gamma} \min \left\{\rho \frac{\log m}{\delta^2}, m^{1/2 - \gamma}, k\right\}\]
iterations to compute a $(1+\delta)$-approximate solution.
\end{theorem}
In their definition, a packing input consists of a matrix $A \in \mathbb{R}^{m \times k}$, a vector $b \in \mathbb{R}^m_{> 0}$, and a polyhedron $P \subseteq \mathbb{R}^k$, such that $Ax \geq 0$ for all $x \in P$. A $(1+\delta)$-approximate solution is an element $x \in P$ that satisfies $Ax \leq (1+\delta) b$. A Dantzig-Wolfe-type algorithm is an algorithm that interacts with $P$ via an oracle $X_P$ that for an input vector $y \in \mathbb{R}^m_{\geq 0}$ can return a solution $x^* \in \argmin_{x \in P} \inp{y}{x}$. As already mentioned, this is a special class of instances for the general resource sharing problem. \\
We will only use a simple consequence in the following: Namely, that for $\gamma \in (0,1/2)$ there exist constants $c_{\gamma},K_{\gamma} > 0$, such that for every $m > K_{\gamma}$ there exists an instance $\mathcal{X}_1 \subseteq \mathbb{R}^m_{\geq 0}$ of the resource sharing problem, with $\lambda^* = 1$ and constant width, i.e. $\max_{x \in \mathcal{X}_1} \norm{x}_{\infty} = \rho_{\gamma} \in \mathcal{O}(1)$, such that every deterministic Dantzig-Wolfe-type algorithm requires $c_{\gamma} \min \left\{ \frac{\log m}{\delta^2}, m^{1/2-\gamma}\right\}$ oracle calls to compute a $(1+\delta)$-approximate solution. \\
Note that these are lower bounds on the number of vertices that lie in the support of the convex combination describing a $(1+\delta)$-approximate solution. So they hold for any Dantzig-Wolfe-type algorithm, independently of the chosen prices. It is also clear that such methods cannot prove stronger lower bounds than $\Omega(m)$ on the number of required oracle calls by Caratheodory's theorem. Therefore, it remains open whether an algorithm with $\tilde{\mathcal{O}}(\frac{m}{\delta^{\alpha}})$ oracle calls for $\alpha < 2$ can exist.\\
We prove that for any version of \cref{alg:generalized_resource_sharing_core} the asymptotic number of oracle calls of our algorithm is best possible in general (for a range of parameters).
\begin{proof}[of \cref{theo:lower_bound_iterations_rs}]
We combine the example above with the scaled probability simplex. For fixed $\gamma \in (0,1/2)$ let $\mathcal{X}_1$ be the instance with width $\rho_{\gamma} \in \mathcal{O}(1)$ (and optimum objective value $\min_{x_1 \in \mathcal{X}_1} \norm{x_1}_{\infty} = 1$), such that every Dantzig-Wolfe-type algorithm requires $c_{\gamma} \min \left\{\frac{\log m}{\delta^2} , m^{1/2-\gamma} \right\}$ oracle calls to compute a $(1+\delta)$-approximate solution. Let $\mathcal{X}_2 := m \Delta_m$ be the scaled probability simplex. We define an instance of the resource sharing problem with two customers $\mathcal{C}=\{1,2\}$. For $0 \in \mathbb{R}^m$, we define the set of feasible resource allocations as $X_1 := \mathcal{X}_1 \times \{0\}$ and $X_2 :=  \{0\} \times \mathcal{X}_2$. Then we set $\mathcal{X} := X_1 + X_2 = \mathcal{X}_1 \times \mathcal{X}_2 \subseteq \mathbb{R}^{2m}_{\geq 0}$. We have $\lambda^* = 1 = \min_{x_1 \in \mathcal{X}_1} \norm{x_1}_{\infty} = \min_{x_2 \in \mathcal{X}_2} \norm{x_2}_{\infty}$. So, a $(1+\delta)$-approximate solution $x =(x_1,x_2) \in \mathcal{X}$ with $x_1 \in \mathcal{X}_1$ and $x_2 \in \mathcal{X}_2$ decomposes into two $(1+\delta)$-approximate solutions in the sets $\mathcal{X}_1$ and $\mathcal{X}_2$. Now let us analyze \cref{alg:generalized_resource_sharing_core}. In every phase, the oracle function of the first customer is called at most $\rho_{\gamma} +1$ times, because $\mathcal{X}_1$ has width bounded by $\rho_{\gamma}$ and thus apart from the last oracle call in the phase it always holds $\xi \geq \frac{1}{\rho_{\gamma}}$. Thus, by construction, we need $T \geq  \frac{c_{\gamma}} {\rho_{\gamma} +1} \min \left\{\frac{\log m}{\delta^2} , m^{1/2-\gamma} \right\}$ to compute a $(1+\delta)$-approximate solution in $X_1$. On the other hand, because we can always assume that the oracle function of the second customer returns a scaled unit vector $me_i$, implying $\xi \leq\frac{1}{m}$. Thus in every phase we have $m$ oracle calls for the second customer. This leads to a total number of oracle calls of $\geq T m$. One can introduce an additional number of $Tn$ oracle calls by adding $n$ customers with feasible regions $\mathcal{X}_3 = ... = \mathcal{X}_{n+2} = \{0\}$. Therefore, any version of \cref{alg:generalized_resource_sharing_core} that computes a $(1+\delta)$-approximate solution uses, for $\tau_{\gamma} := \frac{c_{\gamma}}{\rho_{\gamma}+1}$,  $T(n+m) \geq (n+m) \tau_{\gamma} \min \left\{\frac{\log m}{\delta^2} , m^{1/2-\gamma} \right\}$ many oracle calls.
\end{proof}

\section{Conclusion}
In this work, we have presented an FPTAS for the primal and the dual of the resource sharing problem and improved on the best-known running time in terms of oracle calls. In the last section, we have shown that this number of oracle calls is best possible in a certain sense. Further improvements, if possible, require different types of algorithms. A mere change of the price update rule is not enough to achieve faster convergence. This also implied that no warm-start analysis of the algorithm is possible.\\
Moreover, we were able to show that our algorithm has the natural property to optimize (close to) independent parts of the instance separately by introducing the notion of local weak duality. This implied that our algorithm computes solutions that are close to decreasingly minimal on the two largest entries. Local weak duality provides a theoretical understanding of the empirically observed resilience to local effects of our algorithm. Extending the algorithm to achieve (approximate) decreasing minimality on the third-highest entry and beyond is subject to future work.
\paragraph{Acknowledgments.} We thank Jens Vygen for many fruitful discussions and Sebastian Pokutta for helpful suggestions.

%
%
%
%

\end{document}